\documentclass{aa}
\usepackage{graphicx} 
\usepackage{natbib}
\usepackage{txfonts}
\bibpunct{(}{)}{;}{a}{}{,}
\begin{document}
%
%
\title{Multi-periodic photospheric pulsations and connected 
  wind structures in \object{HD\,64760}
\thanks{Based on observations collected at the European Southern 
  Observatory at La Silla (Proposal IDs 56.D-0235 and 70.D-0433).}}

%
%
\author{ A.~Kaufer\inst{1} \and O.~Stahl\inst{2} \and 
         R.K.~Prinja\inst{3} \and D.~Witherick\inst{3}\fnmsep
         \thanks{\emph{Present address:} Physics Department,
         Trinity College Dublin, Dublin 2, Ireland}}
\institute{
  European Southern Observatory, 
  Alonso de Cordova 3107, 
  Casilla 19001, 
  Santiago 19, Chile
  \and
  Landessternwarte Heidelberg,
  K\"onigstuhl 12, 
  69117 Heidelberg, Germany
  \and
  Department of Physics \& Astronomy, 
  University College London, 
  Gower Street, London, 
  WC1E 6BT, UK
}
\offprints{A.Kaufer, \email{akaufer@eso.org}}
\date{Received ---; accepted ---}
\abstract{ We report on the results of an extended optical
  spectroscopic monitoring campaign on the early-type B supergiant
  \object{HD\,64760} (B0.5\,Ib) designed to probe the deep-seated
  origin of spatial wind structure in massive stars. This new study
  is based on high-resolution echelle spectra obtained with the {\sc
  Feros} instrument at ESO La Silla. 279 spectra were collected over
  10 nights between February 14 and 24, 2003.
  From the period analysis of the line-profile variability of the
  photospheric lines we identify three closely spaced periods around
  4.810\,hrs and a splitting of $\pm3$\%. The velocity -- phase
  diagrams of the line-profile variations for the distinct periods
  reveal characteristic prograde non-radial pulsation patterns of
  high order corresponding to pulsation modes with $l$ and $m$ in the
  range $6-10$. 
  A detailed modeling of the multi-periodic non-radial
  pulsations with the {\sc Bruce} and {\sc Kylie} pulsation-model
  codes \citep{1997MNRAS.284..839T} favors either three modes with
  $l=-m$ and $l=8,6,8$ or $m=-6$ and $l=8,6,10$ with the second case
  maintaining the closely spaced periods in the co-rotating frame.
  The pulsation models predict photometric variations of
  $0.012-0.020$\,mag consistent with the non-detection of any of the
  spectroscopic periods by photometry.
  The three pulsation modes have periods clearly shorter than the
  characteristic pulsation time scale and show small horizontal
  velocity fields and hence are identified as p-modes.
  The beating of the three pulsation modes leads to a retrograde beat
  pattern with two regions of constructive interference diametrically
  opposite on the stellar surface and a beat period of 162.8\,hrs
  (6.8\,days). This beat pattern is directly observed in the
  spectroscopic time series of the photospheric lines. The wind-sensitive
  lines display features of enhanced emission, which appear to follow 
  the maxima of the photospheric beat pattern.
  The pulsation models predict for the two regions normalized flux
  amplitudes of $A=+0.33,-0.28$, sufficiently large to raise spiral
  co-rotating interaction regions \citep{1996ApJ...462..469C}.
  We further investigate the observed H$\alpha$ wind-profile
  variations with a simple rotating wind model with wind-density
  modulations to simulate the effect of possible streak lines
  originating from the localized surface spots created by the NRP beat
  pattern. It is found that such a simple scenario can explain the
  time scales and some but not all characteristics of the observed
  H$\alpha$ line-profile variations.
}
\maketitle
%
%

\section{Introduction}

The discovery of systematic, patterned variability in the stellar
winds of luminous hot stars rates as one of the major landmarks in
recent stellar astrophysics. It is pertinent to our understanding of
wind dynamics and stellar structure.

Clearly, to make progress understanding these new phenomena and their
impact on the fundamental process of mass-loss via stellar winds --
and therefore on stellar evolution scenarios -- , the mechanism(s)
responsible for changing the starting conditions of the wind and --
more specifically -- the mechanisms for dividing the stellar surface
into distinct regions must be determined.  Furthermore, the underlying
physical mechanisms for coupling the photospheric and wind variations
have to be worked out in detail.

The discovery of strictly periodic and sinusoidal modulations of the
\ion{Si}{iii}, \ion{Si}{iv}, and \ion{N}{v} UV resonance lines with
periods of $1.2$ and $2.4$ days \citep{1995ApJ...452L..61P,
1997A&A...327..699F} and the recent detection of the same periods in
the optical H$\alpha$ line \citep[][ KPS\,2002 in the
following]{2002A&A...382.1032K} have turned the intrinsically fast
rotating early-type B supergiant \object{HD\,64760} (B0.5\,Ib) into a
key object in the study of spatially structured hot star winds and
their connection to the stellar surface:

KPS\,2002 presented the first observational evidence for the physical
link between the observed photospheric variations and the large scale
wind structure in \object{HD\,64760}. Photospheric low-order NRP
patterns and the constructive or destructive interference of
multi-periodic photospheric oscillations were proposed as the most
probable source to provide the large-scale perturbations at the base
of the wind needed to build large-scale wind structure.  The
photospheric UV lines of \object{HD\,64760} as observed intensively
with {\sc Iue} were used to probe for the first time the transition
zone between the deeper photosphere and the base of the stellar wind
by \citet{1998MNRAS.296..949H}. Unfortunately, the study was limited
by the comparatively low $S/N$ of the {\sc Iue} spectra. However, this
missing link photosphere -- wind can now be filled observationally
with highest quality \emph{optical} spectroscopic time series, which
is the study presented here.

The optical variability analysis of \object{HD\,64760} by KPS\,2002 was
based on {\sc Heros} time series obtained at the ESO\,50-cm telescope
and suffered from the comparatively low $S/N$ ($\approx 100$ at
H$\alpha$ and dramatically dropping towards the blue) and the poor
time resolution (1 spectrum per night) of the data set. In particular,
the low quality of the spectra inhibited a detailed study of 
\emph{photospheric} line-profile variability, which was found to be very
subtle, i.e., line-profile variations with peak-to-peak amplitudes of
less than $0.5$\% of the continuum level.

In this work, we present the results from a follow-up observation
campaign with {\sc Feros} at the ESO/MPG\,2.2-m telescope at La Silla,
which provided us with continuous high-quality optical spectroscopic
time series over more than two rotational cycles ($\approx10$\,days)
with prerequisite high $S/N$ ($\approx 400$) and spectral
($\approx50\,000$) and time ($\approx10$\,min) resolution.

The new data set was acquired with the aim to detect and to determine
the characteristics of the photospheric and wind line-profile
variations and to further explore the nature of the mechanism that
couples them together.
The scope of this paper is not to carry out detailed modeling of the
observations but to provide detailed input for unified hydrodynamic
stellar wind models, which ultimately need to include pulsation (or
other physical mechanisms like magnetic fields) as the \emph{driver}
for large scale stellar wind structure.

%
%
\section{Observations}

The new observations were carried out with the high-resolution echelle
spectrograph {\sc Feros} \citep{2000SPIE.4008..459K} at the ESO/MPG
2.2-m telescope at La Silla.  A total of 279 spectra with a resolving
power of $R=48\,000$ and a wavelength coverage from $3600-9200$\,\AA\ 
were collected over 10 nights between February 14 and 24, 2003. In a
typical night, some 30 spectra were recorded over time intervals of
$\sim$4\,hours. Table~\ref{tab:obslog} provides a more detailed
record on the distribution of the individual observations over the
different nights. The average exposure time per spectrum was 400\,s
but varied between 200\,s and 820\,s to account for the variations of
the observing conditions. A very high average $S/N$-ratio of $390$ per
spectrum (at 5400\,\AA) was obtained for this time series.

\begin{table}
  \caption{Observation log of the {\sc Feros} observations
    of \object{HD\,64760} at the ESO/MPG 2.2-m telescope.}
  \label{tab:obslog}
  \centering
  \begin{tabular}[h]{cccc} \hline\hline
    Night \# & Spectra \# & UTC range & Date     \\ 
    \hline
    1 &   1 --  31 & 00:38 -- 04:53  & 2003-02-15\\
    2 &  32 --  64 & 00:30 -- 04:51  & 2003-02-16\\
    3 &  65 -- 110 & 00:23 -- 07:47  & 2003-02-17\\
    4 & 111 -- 145 & 00:21 -- 04:56  & 2003-02-18\\
    5 & 146 -- 172 & 00:41 -- 04:54  & 2003-02-19\\
    6 & 173 -- 200 & 00:27 -- 04:54  & 2003-02-20\\
    7 & 201 -- 204 & 00:14 -- 00:49  & 2003-02-21\\
    8 & 205 -- 224 & 00:18 -- 04:44  & 2003-02-22\\
    9 & 225 -- 249 & 00:08 -- 04:43  & 2003-02-23\\
   10 & 250 -- 279 & 00:31 -- 04:47  & 2003-02-24\\
   \hline
  \end{tabular}
\end{table}

Flatfield and wavelength-calibration exposures have been obtained with
the instrument-internal Halogen and Thorium-Argon lamps at the
beginning of the respective nights. All spectra have been reduced
semi-automatically with ESO-{\sc Midas} using the dedicated {\sc Feros} 
context as described e.g.~in \citet{1999ASPC..188..331S}.
All spectra have been reduced to barycentric velocities and have been
normalized to the stellar continuum using the very stable instrument
response curve and low-order fits to clean stellar continuum points.

Throughout this paper, all velocities are given with respect to the
laboratory wavelengths of the corresponding lines of interest. For
this purpose the wavelengths of the respective lines have been
corrected by a systemic velocity of $v_\mathrm{sys} =
+18$\,km\,s$^{-1}$ for \object{HD\,64760} as was obtained from the
fitting of rotationally broadened synthetic line profiles by
\,KPS\,2002 (cf. their Fig.~1). The uncertainty in $v_\mathrm{sys}$
from the fits amounts to $\pm2$\,km\,s$^{-1}$.

\begin{figure}
\resizebox{\hsize}{!}{\includegraphics[angle=0]{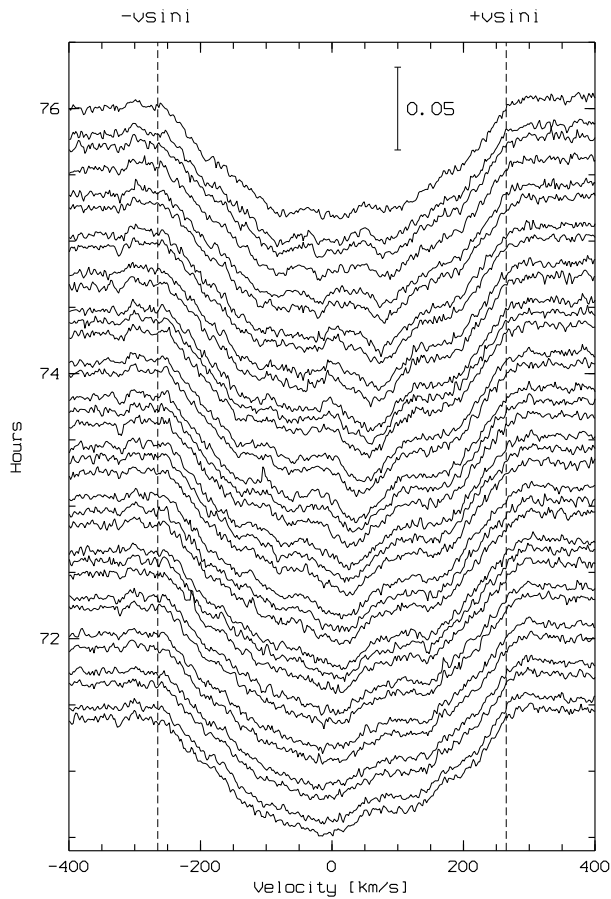}}
\caption{Time series of spectra of the \ion{Si}{iii}$\lambda$4553 line
  in night 4 of our monitoring campaign. The time axis increases 
  towards the top and is given in hours since MJD~$=52685.0265$. The
  vertical bar indicates $5$\% in intensity with respect to the
  continuum; $v\sin i = \pm 265$\,km\,s$^{-1}$ is indicated by
  vertical dashed lines. }
\label{fig:Si4553_night4}
\end{figure}

Figure\,\ref{fig:Si4553_night4} shows an example of the 35 spectra
around the \ion{Si}{iii}$\lambda$4553 line as obtained in night\,4 of
the monitoring campaign. Systematic line-profile variability (LPV) is
directly discernible and will be discussed in detail below.

%
%
\section{Spectroscopic time series }
\label{sect:timeseries}

\begin{figure}
\resizebox{\hsize}{!}{\includegraphics[angle=0]{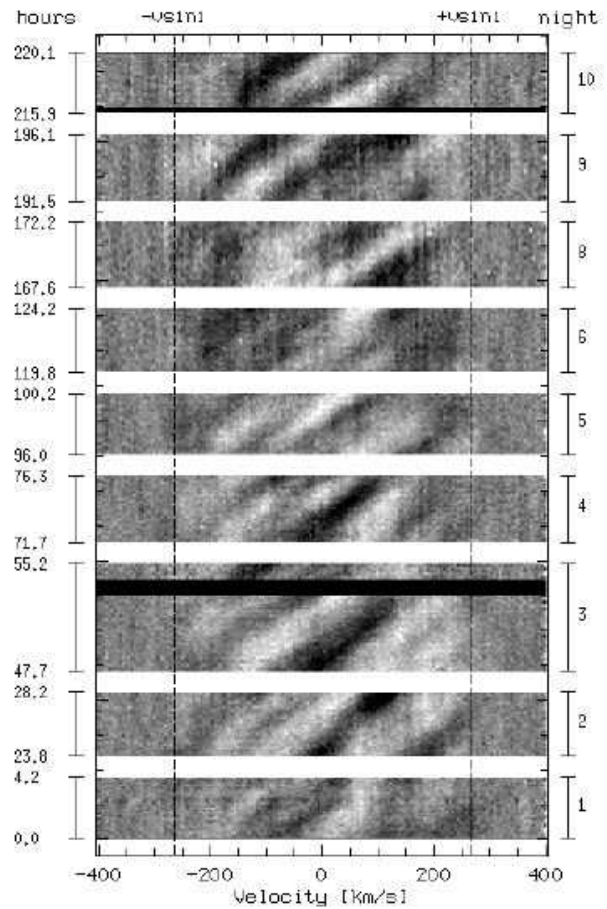}}
\caption{Dynamical velocity -- time spectrum of the
  \ion{Si}{iii}$\lambda$4553 line.  On the right axis the observing
  night and on the left axis the start and end time of the
  corresponding observations is indicated in hours since
  MJD~$=52685.0265$.  The intensities from black to white are displayed
  with $\pm1$\% cut levels. Further, $v\sin i$ is indicated by
  vertical dashed lines.}
\label{fig:Si4553_ts}
\end{figure}

\begin{figure}
\resizebox{\hsize}{!}{\includegraphics[angle=0]{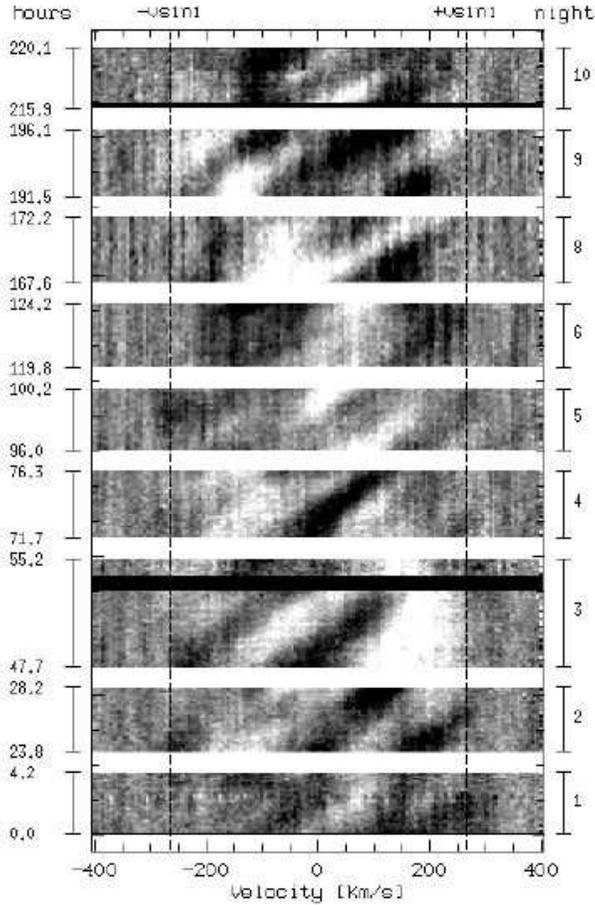}}
\caption{Dynamical spectrum of the \ion{He}{i}$\lambda$6678 line.
  Displayed with $\pm1$\% cut levels. $v\sin i$ is indicated by
  vertical dashed lines.}
\label{fig:He6678_ts}
\end{figure}

\begin{figure}
\resizebox{\hsize}{!}{\includegraphics[angle=0]{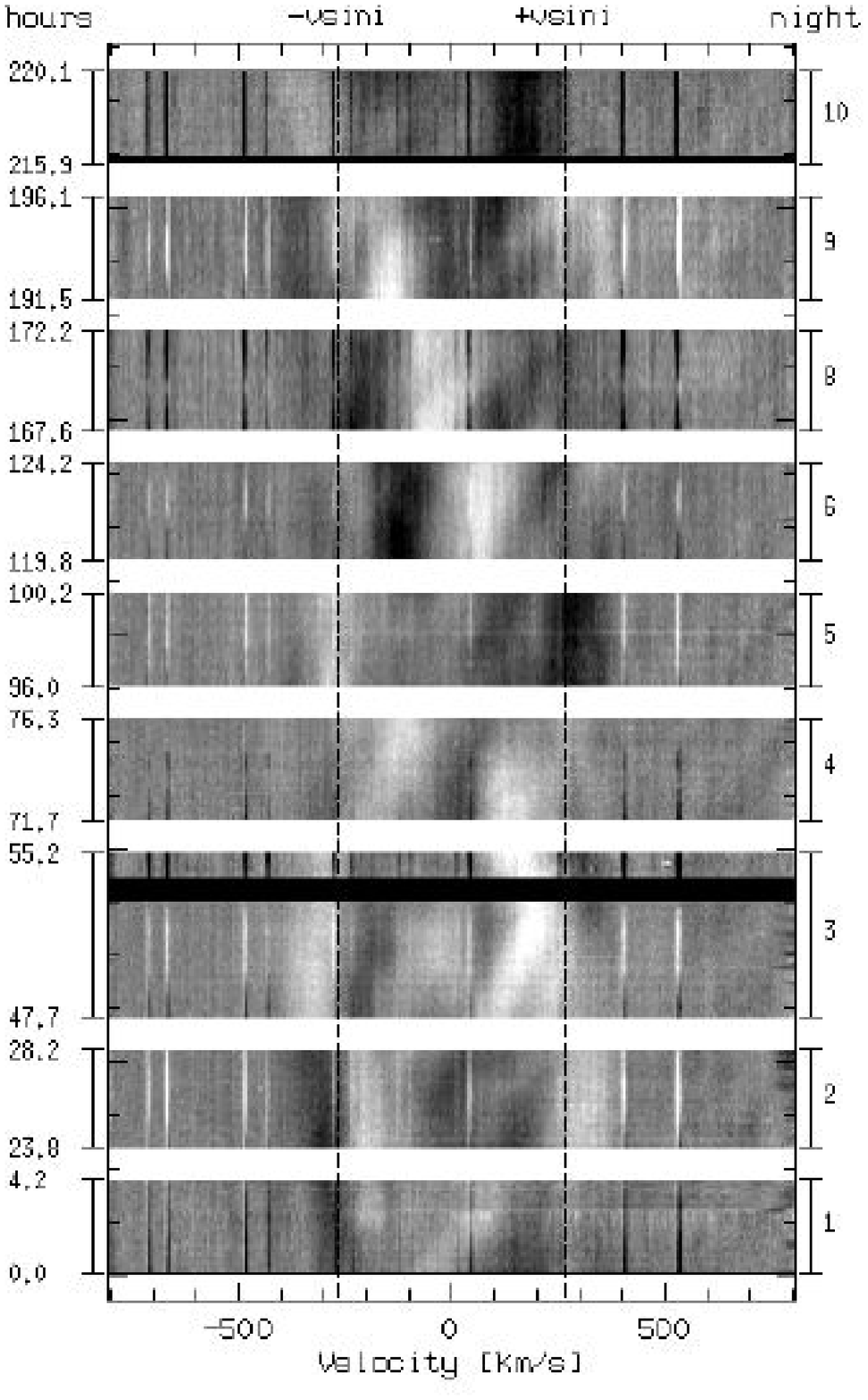}}
\caption{Dynamical spectrum of the H$\alpha$ line. Displayed with
  $\pm3$\% cut levels. $v\sin i$ is indicated by vertical dashed
  lines. Note the extended velocity range of $\pm800$\,km\,s$^{-1}$ to
  cover the wind features of the line profile. The sharp spectral
  features are unresolved telluric lines in the H$\alpha$ region. }
\label{fig:Halpha_ts}
\end{figure}

In the following we present the spectroscopic times series as
so-called dynamic spectra where the time series of spectra is
displayed in a two-dimensional velocity (wavelength) -- time image
with the intensity levels of the individual spectra represented by a
grayscale from the low (black) to the high (white) intensity cut
level. The intensity levels are computed as the \emph{difference}
with respect to the average spectrum of the times series. Therefore,
the discernible absorption (black) and emission (white) features in
the dynamical spectra are pseudo-absorption and pseudo-emission
features relative to the time-averaged stellar spectrum (gray). The
average stellar spectrum is constructed with the weights of the
individual spectra according to their respective $S/N$-ratio.

In this paper, we present the time series in a collapsed form, not
showing the full time gap from night to night. The time gap remains as
a white horizontal bar, while times with no data within the dynamical
spectra of one night appear as black horizontal bars. The seventh
night contains only four spectra and is completely omitted from the
dynamical spectra (but not from the analyses below). The dynamical
spectra indicate on the right axis the observing night and on the left
the start and end time of the corresponding observations as counted in
hours from the start time of the exposure of the first spectrum of the
time series, i.e., $t= 0$\,hrs equals to MJD~$=52685.0265$. In
addition, all dynamical spectra indicate the projected rotational
velocity of $v\sin i = \pm265$\,km\,s$^{-1}$ as was determined in
KPS\,2002 for \object{HD\,64760}.

The dynamical spectra of the representative purely photospheric
\ion{Si}{iii}$\lambda$4553 line, the photospheric
\ion{He}{i}$\lambda$6678 line, which shows some wind emission
contribution and the mostly wind-sensitive H$\alpha$ line are shown in
Figs.~\ref{fig:Si4553_ts}, \ref{fig:He6678_ts} and
\ref{fig:Halpha_ts}, respectively.  Virtually all available isolated
photospheric lines in the spectrum of \object{HD\,64760} in the
accessible wavelength range from 3600-9200\,\AA\ have been examined
through their dynamical spectra and are found to appear very similar
--- but very different from H$\alpha$. The variations in the
photospheric lines are largest in the shown \ion{Si}{iii}$\lambda$4553
and \ion{He}{i}$\lambda$6678 lines with peak-to-peak amplitudes of
$\pm1$\% relative to the time-averaged spectrum.  The strong
\ion{He}{i}$\lambda$6678 line does show in addition to the
photospheric patterns some wind features from H$\alpha$, e.g., most
prominent, a broad pseudo-emission component traveling from the red
(day 3) to the blue side (days 6-9) of the line. The wind sensitivity
of the \ion{He}{i}$\lambda$6678 line is substantiated by a small
amount of asymmetry in the time-averaged line profile and the
existence of small emission peaks superimposed to the photospheric
line profile at $-298$\,km\,s$^{-1}$ and $+330$\,km\,s$^{-1}$. The
emission peaks reach a maximum intensity of 0.4\% above the continuum
and are clear mass-loss indicators for a fast rotating wind as was
demonstrated for H$\alpha$ in KPS\,2002.  No such wind features are
found for the \ion{Si}{iii}$\lambda$4553 line supporting its
\emph{bona fide} photospheric origin.

The photospheric lines vary strongly within one night, but also the
variability pattern looks different from night to night and changes
from periods with low variability (like in nights 1 and 6) to periods
with strong patterns (like in nights 3 and 9). This observed pattern
variability is indicating either a contribution from non-periodic
effects or constructive/destructive multi-periodic effects. The
pseudo-periods become directly discernible if the spectral intensity
variations are followed in time for a fixed velocity bin (e.g. at
zero-velocity) and appear to be of the order of the length of one
observing night (i.e., about 4 to 6 hours). It is important to note
here, that the length of the observing nights was just sufficient to
sample the full extent of most of the photospheric variations as can
e.g. be seen in Figs.~\ref{fig:Si4553_night4} and \ref{fig:Si4553_ts}
in night 4 where the pattern at velocity zero evolves from absorption
to emission and back to absorption within 4.5\,hrs.
At all times, the photospheric lines do clearly show structured
variability with alternating prograde traveling pseudo-absorptions
and pseudo-emission features in the dynamical spectra, which are
indicative of the presence of non-radial pulsations (NRP) in the
atmosphere of \object{HD\,64760}. The fact that the slope of the traveling
features appears to be non-constant and that from night to night
periods of low and high variability amplitudes can be distinguished 
favor the presence of multi-periodic NRPs. 
\citet{1984LIACo..25..115B} reported for the first time photospheric
line-profile variations in highest-quality optical spectra of the
\ion{Si}{iii}$\lambda4553$ and \ion{He}{i}$\lambda6678$ lines, which
where interpreted as high ($|m|$ large) and low order ($|m|\approx2$)
NRPs with respective periods of $\sim0.1$ and $\sim0.5$\,days.
\citet{1998MNRAS.296..949H} could detect the 1.2-day period reported
in the UV wind lines also in the UV photospheric lines of the {\sc Iue
  Mega} data set with a weak signature for a prograde traveling NRP
pattern in the corresponding phase diagrams while KPS\,2002 report a
weak signature of prograde NRPs in the optical \ion{He}{i}$\lambda4026$
line with a 2.4-day period.
Only the high quality and proper sampling over several contiguous
nights as achieved with this new spectroscopic data set allows for the
first time to directly visualize the presence of NRPs in \object{HD\,64760}.
The characterization and modeling of the multi-periodic NRPs in
\object{HD\,64760} are the main subject of our analysis in
Sects.~\ref{sect:periodanalysis} and \ref{sect:nrp_modeling}.

In contrast to the photospheric lines the dynamical spectrum of the
H$\alpha$ line as shown in Fig.~\ref{fig:Halpha_ts} varies very little
within one night, indicating that predominantly longer time scales of
the order of days are governing the variability of this line.  This
finding is consistent with the observations in KPS\,2002. However, so
far no evidence could be given for any variability on shorter time
scales due to the inappropriate 1-day sampling rate in the earlier
{\sc Heros} data set used in KPS\,2002. The new {\sc Feros} data set
provides for the first time high time resolution for the H$\alpha$
line together with coverage over several nights. 
The night-to-night variations in H$\alpha$ show an evolution of the
pseudo-absorption and emission features. At the same time the
variations within one night do not reveal the presence of any shorter
time scales but are consistent with the line-profile changes as
expected from the night-to-night trends. Interestingly an absence of
pseudo-absorptions and emissions is noticeable in the nights 1 and 5
almost coincident with the low amplitudes of variability in the
photospheric lines in the nights 1 and 6.
The H$\alpha$ line of \object{HD\,64760} is strongly affected by the
mass loss of the supergiant star and the rotation of the wind as was
demonstrated in KPS\,2002 (cf. their Fig.~2). However, since their
estimate of mass-loss rate of $9\times10^{-7}$\,M$_\odot$/yr is still
moderate, it is anticipated to find contributions of the underlying
photospheric profile and its variability on the shorter time scales as
found for all other photospheric lines (cf. above). This is not the
case and it is worth noting here, that the absence of NRP patterns in
the dynamical spectra of H$\alpha$ will be 'naturally' explained by
the NRP modeling as carried out in Sect.~\ref{sect:nrp_modeling}.

%
%
\section{Period analysis}
\label{sect:periodanalysis}

\subsection{Photospheric lines}

\begin{figure}
\resizebox{\hsize}{!}{\includegraphics[angle=0]{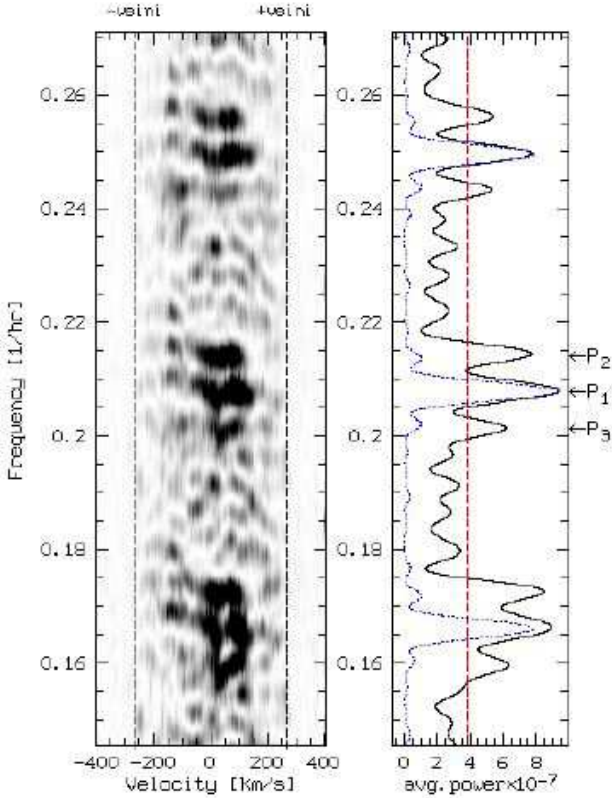}}
\caption{Two-dimensional power spectrum of the
  \ion{Si}{iii}$\lambda$4553 line around a period of 4.8\,hr showing
  the three significant power peaks as identified by the iterative
  {\sc Clean} process. The 99.5\% level of significance of the
  periodogram is indicated by a (red) dashed line. The window function
  of the data set is shown as (blue) dotted line and was shifted to
  the position and normalized to the height of the strongest power
  peak. The strongest alias periods can be seen at 4\,hr and 6\,hr.
  The three significant periods $P_1=4.810$\,hr, $P_2=4.672$\,hr,
  $P_3=4.967$\,hr are labeled in the right plot, which displays the
  average power over the velocity bin of $-v\sin i$ to $+v\sin i$.  }
\label{fig:Si4553power2d}
\end{figure}

\begin{figure*}
  \centering
  \includegraphics[width=17cm]{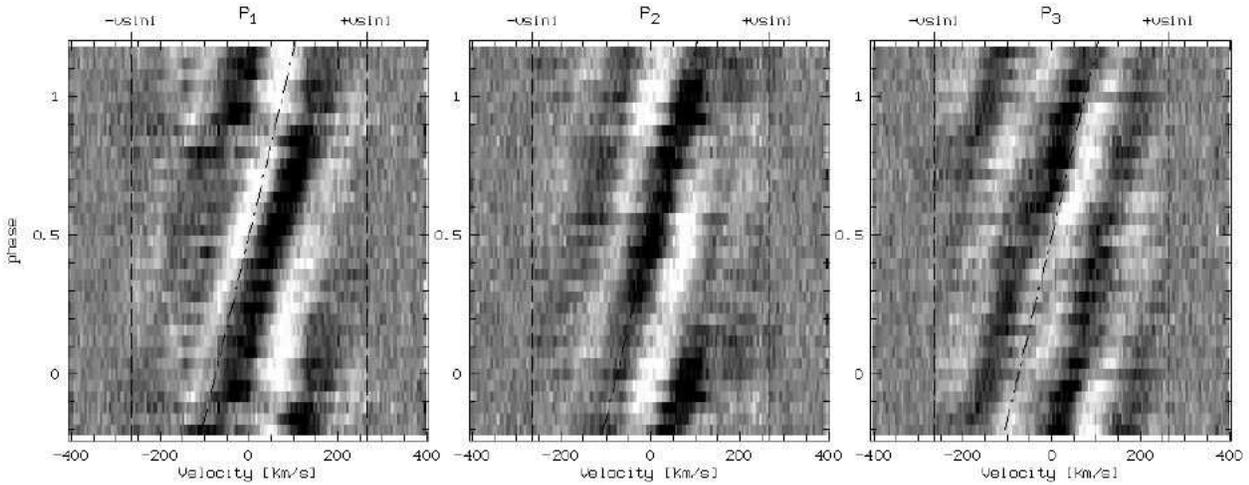}
\caption{Dynamical phase spectra of the \ion{Si}{iii}$\lambda$4553
  line. The observed spectra are phased with $P_1=4.810$\,hr (left),
  $P_2=4.672$\,hr (center), $P_3=4.967$\,hr (right). 25 phase bins
  were used for the phase interval $0.0 - 1.0$. All spectra are
  displayed with $\pm0.3$\% cut levels. The measured typical
  acceleration of the features over the line center of $(dv/d\phi) =
  -175$\,km\,s$^{-1}$/cycle is indicated by a dot-dashed lines.}
\label{fig:Si4553_p123}
\end{figure*}

The qualitative findings from the inspection of the dynamical
spectra of photospheric and wind lines in Sect.~\ref{sect:timeseries}
will be quantified in this section by means of a rigorous time-series 
analysis. 

As a first step we carried out a two-dimensional period analysis on
the {\sc Feros} data set by computing power spectra and periodograms
for velocity bins of $5$\,km\,s$^{-1}$ width across the lines of
interest. The full frequency range sampled by the data set from
$1/10$\,days$^{-1}$ to $1/10$\,min$^{-1}$ was searched for significant
periods by application of the {\sc Clean} algorithm described by
\citet{1996ApJ...460L.107S}.  This algorithm makes use of the
equivalence of the classical {\sc Clean} algorithm and least-square
fits of sine functions to the observed data
\citep{1978A&A....65..345S} --- both performing an iterative
deconvolution of the window function from the time series.
Periodograms and power spectra are computed using Lomb-Scargle
statistics, which take into account for uneven sampling of the data
and have a known exponential $\chi^2(2)$ probability distribution.
Peaks in the periodograms below a level of significance of 99.5\%
probability were rejected from the analysis. For a more detailed
description of this {\sc Clean} implementation
cf.~\citet{1997A&A...320..273K}.

Figure~\ref{fig:Si4553power2d} shows the section of the two-dimensional
power spectrum where the most significant periods are found in the
photospheric lines, i.e., close to a period of 4.8\,hr. Application of
the iterative {\sc Clean} algorithm as described above confirms that
three significant periods with $P_1=4.810$\,hr, $P_2=4.672$\,hr,
$P_3=4.967$\,hr are present. The three periods are sorted by their
associated power and significance with $P_1$ and $P_2$ being of
comparable level but containing about twice as much power than $P_3$.
The power within each detected period is not evenly distributed over 
the velocities but strongly peaked at zero velocity indicating
stronger variability at the line center compared the wings of the
line.
Since the actually measured periods show some variations of the order
of 1.25\% as function of the velocity within the line profile, the
periods given above have been measured from the position of the power
peaks averaged over the full velocity interval $-v\sin i$ to $+v\sin
i$, i.e., $-265$ to $+265$\,km\,s$^{-1}$ (cf.~Fig.~\ref{fig:Si4553power2d},
right plot).

To assure ourselves of the detected multi-periodicity of the
photospheric variations we have to exclude that they are possibly
caused by the complex window function of our observations. This is
implicitly guaranteed by the {\sc Clean} method, which we have used to
identify the periods. However, we have verified the multi-periodicity
in addition as part of the pulsation modeling of the variability,
which will be described in Sect.~\ref{sect:nrp_modeling}.


The period analysis confirms the multi-periodicity of the photospheric
line-profile variations in \object{HD\,64760} as had already been
suspected from Fig.~\ref{fig:Si4553_ts}. It should be noted that the
three periods are closely spaced in time with $\pm3$\% separation and
two periods show similar power, which are the prerequisites for the
possible existence of beat periods on time scales much longer than the
underlying periods themselves.

The two-dimensional velocity -- phase diagrams of the photospheric
\ion{Si}{iii}$\lambda$4553 line constructed with the three identified
periods are shown in Fig.~\ref{fig:Si4553_p123} and clearly show the
characteristic traveling features of three distinct prograde 
NRP modes. While the associated pulsation modes of $P_1$ and $P_3$
appear to be very similar, the pulsation mode of $P_2$ is clearly
different as can be seem from the different slope of the NRP pattern.

Alternatively to the identification of the origin of the line-profile
variations (LPV) as non-radial pulsation patterns, similar LPVs could be
produced by the rotational modulation by regular surface structures.
For $P_1$ a crossing time of an individual surface feature from
$-v\sin i$ to $+v\sin i$ is determined to be 13\,hrs (according to a
measured acceleration of 200\,km\,s$^{-1}/4.81$\,hrs at the line center). 
This would corresponds to a stellar rotation period of $P_\mathrm{rot}
= 26$\,hrs. The breakup rotation velocity for \object{HD\,64760},
however, is estimated to $P_\mathrm{rot,break}=64$\,hrs assuming a
stellar radius of $R=22$\,R$_\odot$ and mass of $M=20$\,M$_\odot$
(cf.\,KPS\,2002).  From this simple estimate it must be concluded that
the observed LPVs cannot be explained by co-rotating surface
structures. Further, the observation of closely spaced multiple
periods is not compatible with rotational modulation where only the
stellar rotation period or integer fractions of the rotation period
can be observed.

Therefore, we have to conclude that the most likely source for the
observed LPVs are multi-periodic non-radial pulsations. In
Sect.~\ref{sect:nrp_modeling} we will further substantiate this
hypothesis by detailed modeling of the NRPs.

\subsection{The wind-sensitive H$\alpha$ line}

\begin{figure*}
  \centering
  \includegraphics[width=17cm]{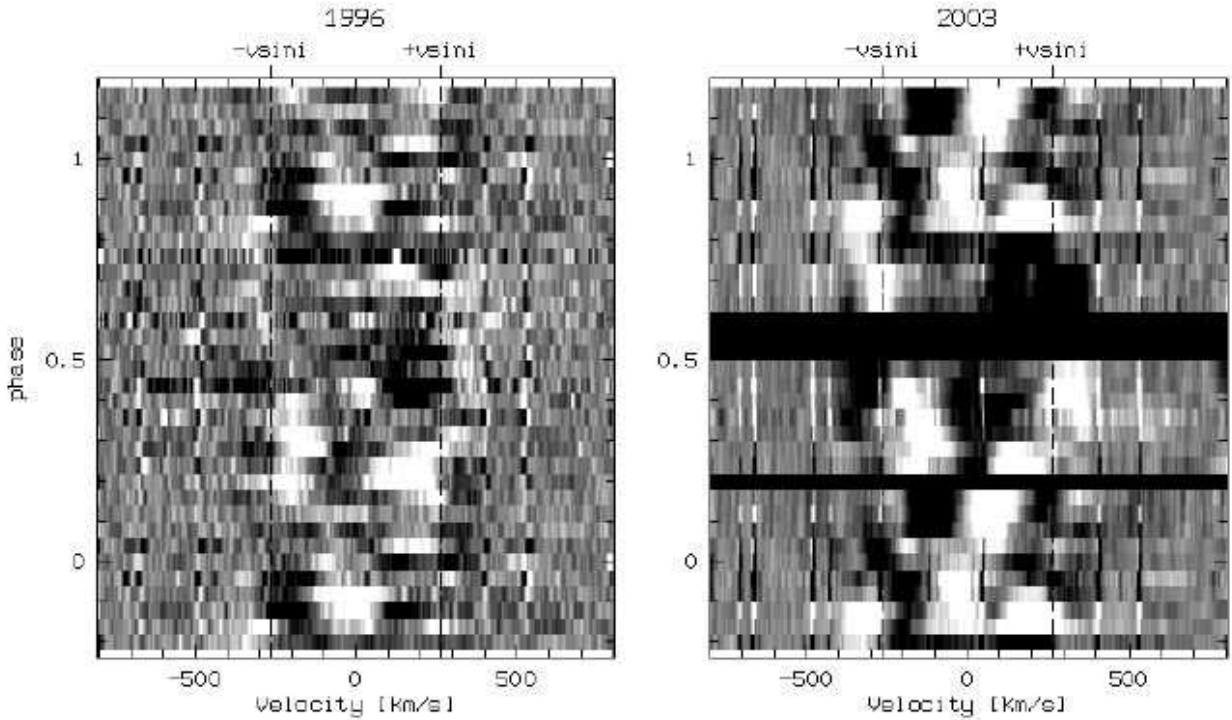}
\caption{Dynamical spectra of the H$\alpha$ line phased with a period
  of 2.39097\,days (cf. text). 25 phase bins were used for the
  phase interval $0.0 - 1.0$. Left 1996 {\sc Heros} data set, right
  2003 {\sc Feros} data set. Both dynamical spectra are displayed with
  $\pm1$\% cut levels.}
\label{fig:Halpha_phase}
\end{figure*}

The H$\alpha$ line has been examined for periodicities in the same way
as the photospheric lines using the iterative {\sc Clean} algorithm
described above. The power spectrum of H$\alpha$ is very complex in
particular in the frequency regime of $<1$\,day$^{-1}$ where it is
hardly possible to identify the most significant frequencies to which
the {\sc Clean} procedure shall be applied. Therefore, we had to closely
inspect selected frequency ranges for which periodicities could be
expected. First, we tried to verify the presence of the three
photospheric pulsation periods close to 4.8\,hrs. But no significant
power could be found as expected from the lack of fast moving features
in the dynamical spectrum of H$\alpha$.

The short extent in time of the new {\sc Feros} data set (10\,nights
with several spectra within a few hours of each night) is not very
well-suited to search for the continued presence of the 2.4-day wind
modulation period as it was found in the {\sc Iue Mega} data set
(15\,days with continuous sampling) and confirmed in H$\alpha$ with
the {\sc Heros} data set (100\,nights with 1 spectrum per night).
Nevertheless, the {\sc Clean}ed two-dimensional power spectra show
significant power at a period of 2.4\,days.  If the H$\alpha$ data
sets of the {\sc Heros} (1996) and the {\sc Feros} (2003) are both
phased with the established period close to 2.4\,days, a
\emph{strikingly similar} variability pattern is found for both
optical data sets as can be seen in Fig.~\ref{fig:Halpha_phase}. Both
data sets have been treated in exactly the same way by sorting the
spectra in 25 phase bins (for phase $0.0 - 1.0$) according to a period
of 2.39097\,days. Phase $0.0$ is defined as the time of the
observation of the first spectrum of the 2003 time series, i.e.,
MJD~$=52685.0265$. The apparent accuracy of the period comes from the
requirement to match the 1996 and the 2003 phase diagrams.  However,
this requirement is fulfilled by a set of discrete periods.  Using
phase-dispersion minimization of the combined 1996+2003 data set, we
have identified 20 such possible periods in the range from $2.37 -
2.41$\,days. Both time series have been resampled to a velocity
resolution of $5$\,km\,s$^{-1}$ and are displayed with $\pm1$\%
intensity cut levels.  The respectively subtracted time-averaged
H$\alpha$ spectra from the 1996 data set (cf.~KPS\,2002, their Fig.~1)
and the 2003 data set to compute the shown residuals are
indistinguishable within the limitations of available $S/N$-ratio and
achievable accuracy in the normalization of the stellar continuum.
The variability pattern of the 2.4-day phased H$\alpha$ line in
Fig.~\ref{fig:Halpha_phase} is very characteristic with 'X' or
'8'-shaped pattern for the pseudo-emission features with the 'crossing
point' close to phase 0.20 (or night 6 in the 2003 data set). 
KPS\,2002 described the 2.4-day H$\alpha$ variability in detail and
tried to disentangle the different contributing components at distinct
velocities in the line profile. All the features highlighted there can
also be identified in the new {\sc Feros} data set --- however, the
features appear to be more coherently traveling across the line
profile.  
The resemblance of the two dynamical H$\alpha$-spectra in 
Fig.~\ref{fig:Halpha_phase} indicates that $2.4$-day period has now
been stable for at least eight years, i.e., the time span between the
1995 {\sc Iue}, the 1996 {\sc Heros} and the 2003 {\sc Feros}
campaigns. Such a long-term presence of a single modulation period
requires a stable clock in the system.  Stellar rotation has been
favored by all previous works on \object{HD\,64760},
cf.~\citet{1997A&A...327..699F} and KPS\,2002. In this work we will
further investigate the possibility that stellar oscillations could
act as stable clock in \object{HD\,64760}, too. This possibility will be
discussed in context with our pulsation and wind modeling in
Sects.~\ref{sect:nrp_modeling} and ~\ref{sect:wind_modeling}.

%
%
\section{NRP line-profile modeling}
\label{sect:nrp_modeling}

The period analysis of the photospheric lines has revealed the
presence of multi-periodic variability with three dominant periods.
We have identified non-radial pulsations (NRPs) as the most likely
origin of this variability and proceed now to confirm this hypothesis
with detailed modeling of the LPVs.

In the following we first concentrate on the determination of the
pulsation quantum numbers $l$ and $m$ of the NRP modes associated with
the three pulsation periods. Only detailed modeling will allow us to
test the consistency of the very complex observed LPVs with
multi-periodic NRPs. However, because of this high complexity of the
observations on one hand side and the very large parameter space on
the other, most likely, an exact identification of the pulsation
parameters is not possible with our observational data set.

\begin{table}
\caption{{\sc Bruce/Kylie} model parameters}
\label{tab:brucekylie_modpara1}
\centering
\begin{tabular}{lrl}\hline\hline
\multicolumn{3}{c}{Model Parameters}    \\
\hline
 $T_\mathrm{pole}$     & 29\,000\,K         & temperature at pole           \\
 $M$                   & 20\,M$_\odot$      & total mass                    \\
 $R_\mathrm{pole}$     & 18\,R$_\odot$      & radius at pole                \\
 $v_\mathrm{equator}$  & 265\,km\,s$^{-1}$  & rotation velocity at equator  \\
 $i$                   & 90\,deg            & inclination                   \\
\hline
\multicolumn{3}{c}{Derived Parameters}  \\
\hline
 $R_\mathrm{equator}$       &  21.6\,R$_\odot$ & radius at equator      \\
 $\log g_\mathrm{pole}$     &  3.23            & gravity at pole        \\
 $\log g_\mathrm{equator}$  &  2.85            & gravity at equator     \\
 $T_\mathrm{equator}$       &  23\,300\,K      & temperature at equator \\
\hline
\multicolumn{3}{c}{Observable Parameters}   \\
\hline
 $P_\mathrm{rot}$  &  98.9\,hr (4.12\,d) & rotation period        \\
 $L_\mathrm{tot}$  &  $1.55\times10^5$\,L$_\odot$   & luminosity  \\
 $T_\mathrm{eff}$  &  24\,600\,K         & effective temperature  \\
\hline
\end{tabular}
\end{table}

We use the NRP modeling codes {\sc Bruce} and {\sc Kylie} (Version
2.84) by \citet{1997MNRAS.284..839T} to examine the LPVs observed in
the photospheric lines of \object{HD\,64760}. {\sc Bruce} calculates the
pulsational perturbations to the rotating stellar surface and includes
pulsational velocity fields in all three spatial directions ($r, \phi,
\theta$), temperature fluctuations, variations in the visible surface
area and the variations in the viewing angle of a given surface
element. {\sc Kylie} derives in a second step the observable
quantities from the perturbation fields provided by {\sc Bruce} by
computing and integrating the observables like the observer-directed
spectrum or flux of each surface element individually from an input
grid of stationary atmosphere models.

The intrinsic line profiles and spectral energy distributions, which
are required as input to {\sc Kylie} for the production of realistic
non-radial pulsation LPVs and light curves, respectively, were
computed on the basis of an extended grid of {\sc Atlas}\,9 model
atmospheres \citep{1993KurCD..13.....K} and subsequent {\sc Bht}
spectrum synthesis \citep{1966AAHam...8...26B} for solar abundances
and a microturbulence of $\xi_\mathrm{micro}=2$\,km\,s$^{-1}$.  The
grid extends from 15\,000 to 31\,000\,K in steps of 1000\,K in
effective temperature $T_\mathrm{eff}$ and from 2.6 to 3.5 in steps of
0.1 in gravity $\log g$.
For the hotter atmospheres (25\,000\,K and higher), the {\sc Atlas}\,9
models with the lowest $\log g$ values do not converge anymore.
Fortunately, these grid points are not required in the {\sc Kylie}
line-profile computations, since for a fast rotating star like
\object{HD\,64760} the hottest regions are located at the stellar pole where
the $\log g$ is high because of the rotational flattening of the star.
The lower temperatures occur close to the equator, where also $\log g$
is found to be lower.

The stellar input parameters for {\sc Bruce} were chosen in such a way
to reproduce the observable parameters of \object{HD\,64760} as given
in KPS\,2002 (their Tab.~1): a polar radius of 18\,R$_\odot$, a polar
temperature of 29\,000\,K, a stellar mass of 20\,M~$_\odot$, an
equatorial rotation velocity of 265\,km\,sec$^{-1}$ (65\% of the
critical rotational velocity of 415\,km\,sec$^{-1}$) and an
inclination of 90$^\circ$. These input values result in an equatorial
radius of 21.6\,R$_\odot$, $\log g$ values of 3.23 and 2.85 at the
pole and equator, respectively, an equatorial temperature of
23\,300~K, and a rotation period of 98.9\,hours (4.12\,days). The
total luminosity and effective temperature, as seen from the observer,
are $1.55\times10^5$\,L$_\odot$ and 24\,600\,K. The model parameters
related to the stationary stellar model are summarized in
Tab.~\ref{tab:brucekylie_modpara1}.

\subsection{Pulsation mode parameters}

For the modeling of  $n$ pulsation modes, {\sc Bruce} needs the 
pulsation periods $P_n$ as measured in the observer's system, the 
pulsation amplitudes $A_n$, the relative pulsation phases $\phi_n$
and the pulsation quantum numbers $m_n$ and $l_n$.

The pulsation quantum number $m$ is equal to the number of meridian
circles on the stellar surface and can be estimated via the
acceleration of the traveling features across the line center. From
the phase diagrams on Fig.~\ref{fig:Si4553_p123} we measure an
acceleration of $(dv/d\phi) = -175\pm50$\,km\,s$^{-1}$/cycle.  This
translates with the purely geometrical relation
$m_n (dv/d\phi)_n/2 = 2 |v\sin i|$ 
into a possible range of values for $m$ of $-8.5< m <-4.7$.

While the $m$-value can be estimated with a reasonable accuracy by
this method, the $l$-value remains ill-defined and can only be
determined by detailed modeling of the LPVs. However, the presence of
multiple pulsation periods and the high but still finite $S/N$-ratio
of the data will naturally limit the achievable accuracy of the
$l$-value determination.

The amplitudes and relative phases of the pulsation modes have been
chosen so that the best possible match to the observations is
achieved: pulsation amplitudes of $3-10$\,km\,s$^{-1}$ are needed to
produce LPVs of the observed amplitudes of $\pm1$\% in the
\ion{Si}{iii}$\lambda$4553 line; the relative phases have been
adjusted to match the times of strongest variations in the observed
time series, i.e., the times of constructive interference of the
pulsation modes.

Our adopted pulsation mode parameters are summarized in
Table~\ref{tab:brucekylie_modpara2}. Two fundamental types of possible
pulsation modes have been investigated, $l=-m$ (Models 1 and 3) and 
$l\neq-m$ (Model 2) and are discussed below.

\begin{table}
\caption{{\sc Bruce/Kylie} model parameters}
\label{tab:brucekylie_modpara2}
\centering
\begin{tabular}[lr]{ccccccc}
\hline\hline
\multicolumn{7}{c}{Pulsation Modes} \\
\hline
$n$ & $P_n$ [hr] & $P_{n,\mathrm{corot}}$ [hr] & 
$\phi_n$ [deg] & $l_n$ & $m_n$ & $A_n$ [km\,s$^{-1}$]\\ 
\hline
\multicolumn{7}{c}{Model 1} \\
\hline
1 & 4.810 & 7.872 & 180 & 8 & $-8$ &  5 \\
2 & 4.672 & 6.520 & 216 & 6 & $-6$ &  5 \\
3 & 4.967 & 8.302 &  97 & 8 & $-8$ &  3 \\
\hline
\multicolumn{7}{c}{Model 2} \\
\hline
1 & 4.810 & 6.791 & 180 &  8 & $-6$ & 10 \\
2 & 4.672 & 6.520 & 216 &  6 & $-6$ & 5 \\
3 & 4.967 & 7.109 &  97 & 10 & $-6$ & 10 \\
\hline
\multicolumn{7}{c}{Model 3} \\
\hline
1 & 4.810 & 7.872 & 180 & 6 & $-6$ &  5 \\
2 & 4.672 & 6.520 & 216 & 8 & $-8$ &  5 \\
3 & 4.967 & 8.302 &  97 & 6 & $-6$ &  3 \\
\hline
\end{tabular}
\end{table}

The measured and derived stellar and pulsation-mode parameters in 
Tabs.~\ref{tab:brucekylie_modpara1} and \ref{tab:brucekylie_modpara2}
allow us to identify the expected pulsation-mode characteristics.
The dimensionless pulsation frequency in the co-rotating frame 
$\omega_\mathrm{corot}$ is defined as 
\begin{eqnarray}
\omega_\mathrm{corot}^2 &=& \frac{(2\pi/P_\mathrm{corot})^2 R_\mathrm{pole}^3}{GM} 
\end{eqnarray}
and reaches values of $\omega_\mathrm{corot}^2\approx30$ for the
observed pulsation parameters of \object{HD\,64760}. This places the
three observed pulsation modes in the regime of the pressure- or
\emph{p-mode} pulsations \citep{1997PhDT........24T}.  This finding
is consistent with the observed concentration of power at the line
center in the two-dimensional power spectrum
(Fig.~\ref{fig:Si4553power2d}) indicating a strong preference of
radial over horizontal velocity fields, which is the most direct
characteristics of p-mode pulsations. In the NRP models, the ratio of
the horizontal to the radial velocity amplitudes $k$ is described as
$k=1/\omega_\mathrm{corot}^2$ and will provide a good representation
of the power distribution over the line profiles (cf.
Fig.~\ref{fig:Si4553mod91power2d} below).

\subsection{The case $l = -m$}

The analysis of the phase diagrams above provides a reasonable
estimate of the $m$-value of each pulsation period but the degree $l$
remains ill-defined. $m$ can assume values from $-l$ to $+l$, i.e.,
the lowest acceptable value of $l$ in any case is $-m$. For a given
choice of $l$, there will be $l-|m|$ nodes along any longitudinal
line, and consequently, the stellar surface will be divided into
$l-|m|+1$ sections in latitude. With an increasing number of zones on
the stellar surface the mutual cancellation of the pulsation effects
of neighboring zones will increase. Therefore, to produce the maximum
observable effects, the smallest number of surface zones should be
chosen, i.e., the case $l=-m$. This is the reason, why in many works
on NRPs, this choice is preferred. However, usually the same
amplitudes in the observables but with different characteristics can
be achieved with higher amplitudes of the modulations in velocity or
temperature in the case $l \neq m$.

\begin{figure}
 \resizebox{\hsize}{!}{\includegraphics[angle=0]{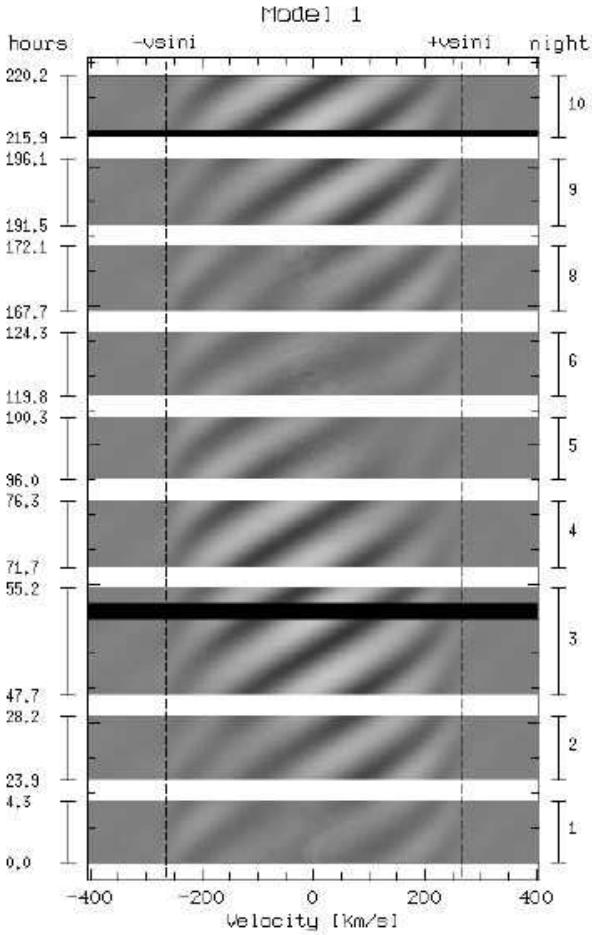}}
 \caption{Computed time series for the \ion{Si}{iii}$\lambda$4553 line
  according to NRP Model~1. Note the strong changes from night to night,
  which are due to the superposition of modes with different $m$
  values. The choice of the $m$-values for the different periods leads
  in this model to a retrograde propagation of the beat pattern -- 
  consistent with the observations in Figs.~\ref{fig:Si4553_ts}
  and \ref{fig:He6678_ts}. Displayed with 1\% cut levels.}
 \label{fig:Si4553_mod91}
\end{figure}

\begin{figure*}
  \centering
  \includegraphics[width=16cm]{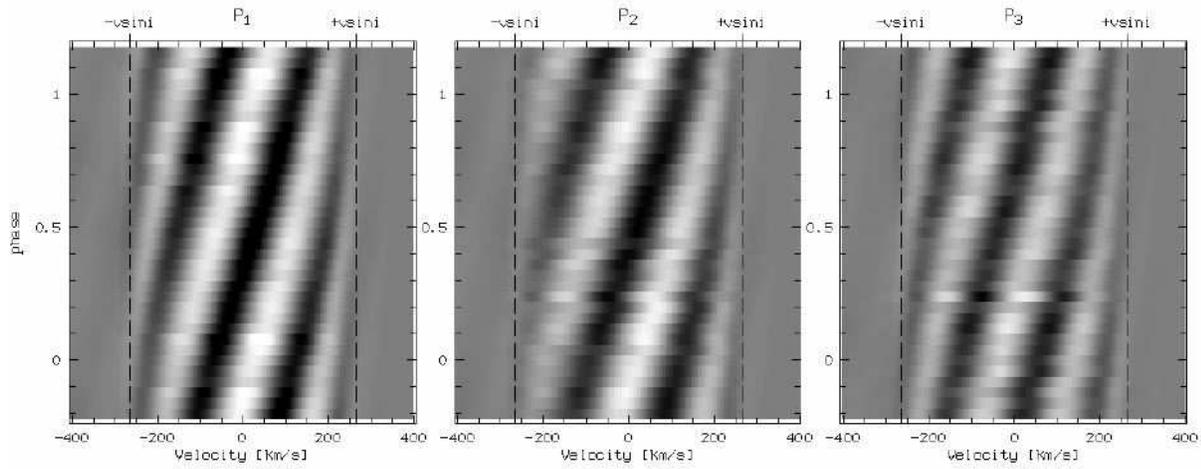}
\caption{Dynamical phase spectra for the the
  \ion{Si}{iii}$\lambda$4553 line as computed for Model~1. The spectra
  are phased with $P_1=4.810$\,hr ($l=8, m=-8$), $P_2=4.672$\,hr
  ($l=6, m=-6$), $P_3=4.967$\,hr ($l=8, m=-8$).  25 phase bins were
  used for the phase range $0.0 - 1.0$. All spectra are displayed with
  $\pm0.3$\% cut levels.}
\label{fig:p123mod91}
\end{figure*}

\begin{figure}
 \resizebox{\hsize}{!}{\includegraphics[angle=0]{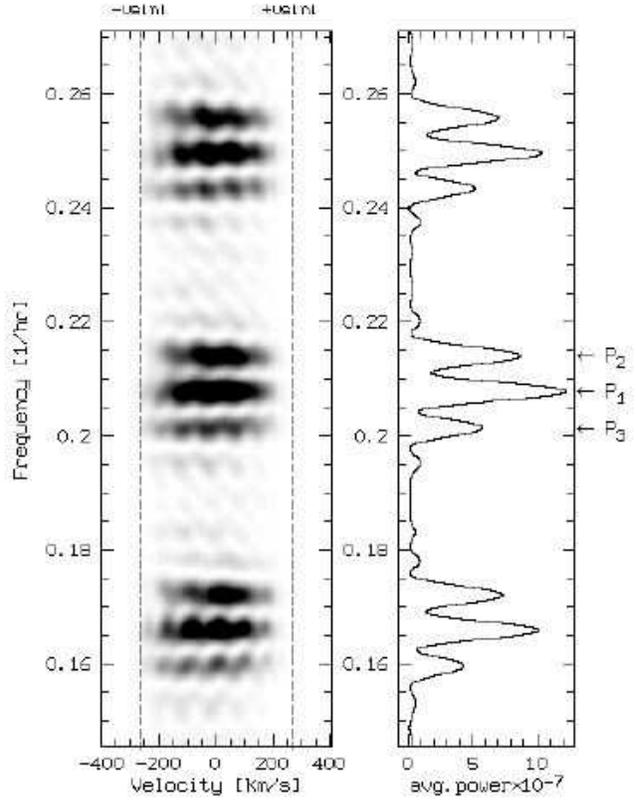}}
  \caption{Two-dimensional power spectrum of the
    \ion{Si}{iii}$\lambda$4553 line around a period of 4.8\,hr for
    Model~1. The power distribution in velocity and frequency of the
    observations (Fig.~\ref{fig:Si4553power2d}) is reproduced very
    well by this model. The concentration of power at the line center
    is due to the p-mode characteristics of the pulsation modes.}
  \label{fig:Si4553mod91power2d}
\end{figure}

\begin{figure}
 \resizebox{\hsize}{!}{\includegraphics[angle=0]{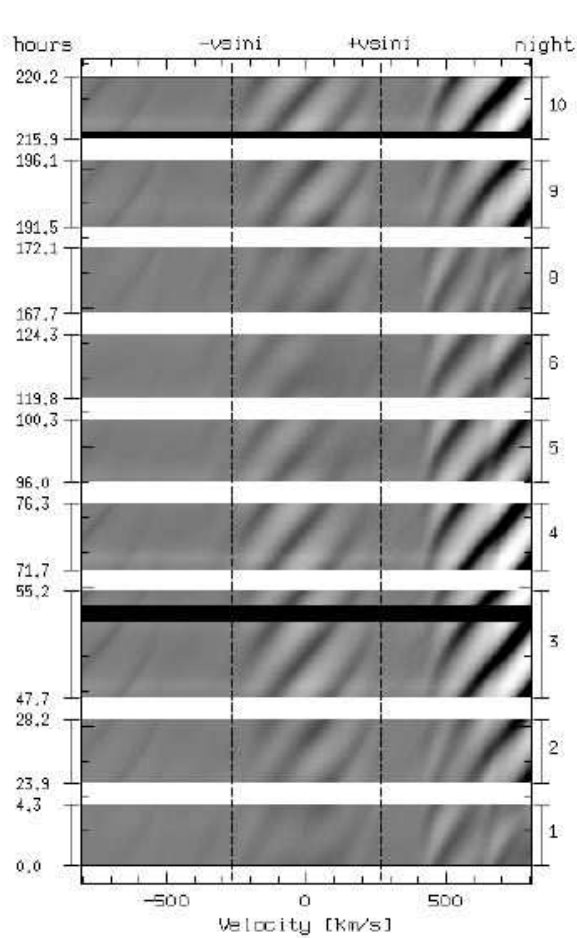}}
 \caption{Computed time series for the photospheric contribution of
   the H$\alpha$ line according to Model~1. The expected variations in
   the line due to NRPs are very small consistent with the
   observations where the pulsation periods could not be detected. The
   strong variations at the right side are due to the neighboring 
   \ion{C}{ii}$\lambda\lambda$6578,6582 lines.  }
 \label{fig:Halpha_mod09}
\end{figure}

\begin{figure}
 \resizebox{\hsize}{!}{\includegraphics[angle=0]{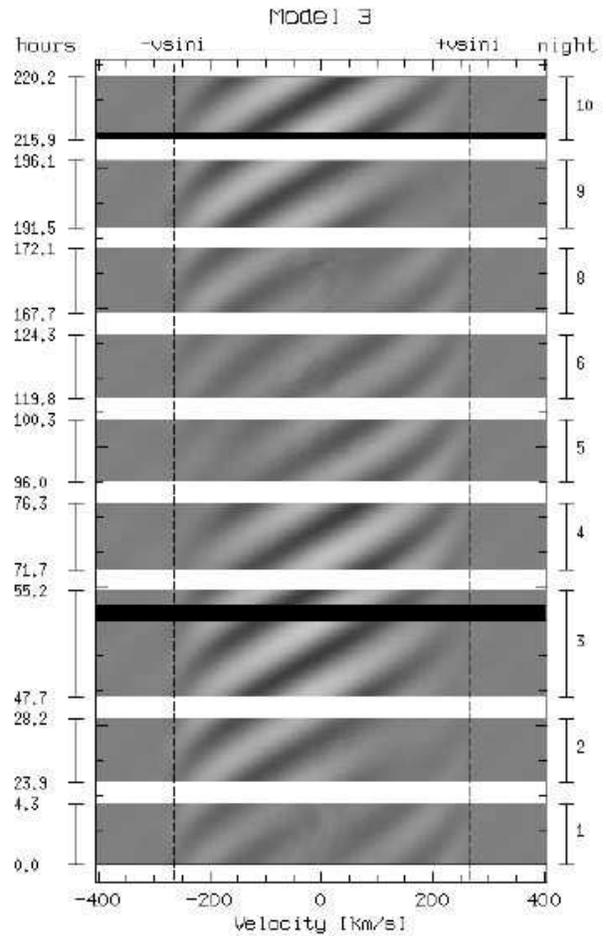}}
 \caption{Computed time series for the \ion{Si}{iii}$\lambda$4553 line
  according to Model~3. The fact that the larger $m$-values has been
  assigned to the longer pulsation period leads to a prograde traveling
  beat pattern, which is not consistent to the observations. Displayed 
  with 1\% cut levels.}
 \label{fig:Si4553_mod90}
\end{figure}

We assume in our first test model (Model\,1) $l = -m$ and $l = 6$ for
the $P_2=4.672$\,hr and $l = 8$ for the $P_1=4.810$\,hr
$P_3=4.968$\,hr periods.  For pulsations with different $m$-values, an
interesting pattern for the resulting total pattern is produced:
several temperature and velocity maxima emerge along the equator from
the constructive or destructive interference of the pulsation modes.
Depending on the period difference of the $m$-modes, this beat pattern
moves across the stellar disk in the direction of the rotation (cf. our 
Model~3, Fig.~\ref{fig:Si4553_mod90}), but also against this direction 
(this Model~1, Fig.~\ref{fig:Si4553_mod91}). The relevant period is the 
beat period of the two modes.

For a best possible comparison with the observations, we computed the
NRP models with exactly the same sampling pattern as during the
observations. Dynamical spectra of the computed time series of the
\ion{Si}{iii}$\lambda$4553 line and the photospheric part of the
H$\alpha$ line for Model~1 are shown in Figs.~\ref{fig:Si4553_mod91}
and \ref{fig:Halpha_mod09}, respectively.  The model for the
\ion{Si}{iii}$\lambda$4553 line reproduces the strong changes from
night to night in the observed variability pattern. While the
dominating non-radial pulsation features travel prograde over the line
profile within a few hours, the constructive interference of the $P_1$
and $P_2$ pulsation modes travels \emph{retrograde} over the line
profile with a period determined by the beat period of the two
dominant pulsation modes:
\begin{eqnarray}
\label{eqn:Pbeat}
 \frac{1}{P_\mathrm{beat}} & = & \frac{1}{P_2} - \frac{1}{P_1} 
    =   \frac{1}{4.672\,\mathrm{hr}} - \frac{1}{4.810\,\mathrm{hr}} 
    =   \frac{1}{162.8\,\mathrm{hr}} 
\end{eqnarray}
The maximum amplitude of the constructive interference is indeed
observed on the nights 3 and 9 corresponding to the beat period of
162.8\,hr or 6.8\,days. Further, the observed phases with low
variability in nights 1 and 6 (cf.~Figs.~\ref{fig:Si4553_ts},
\ref{fig:He6678_ts}) are quite well reproduced by this model with
three beating periods. It is important to note here the directly
observable secondary effects on the stellar wind as seen in the
superimposed wind emission in the \ref{fig:He6678_ts} line and -- even
more pronounced -- in the H$\alpha$ line. The emission feature travels
across the line profile from red to blue and seems to follow the
retrograde pattern and is strongest in the nights 2 to 4 and 6 to 9 .

Figures~\ref{fig:p123mod91} and \ref{fig:Si4553mod91power2d} show the
phase diagrams and the two-dimensional power spectrum as computed from
Model~1 and have to be compared to the observations in
Figs.~\ref{fig:Si4553_p123} and \ref{fig:Si4553power2d}, respectively.
An excellent agreement with the observations is achieved with this
multi-periodic NRP model, including the p-mode characteristics of
the pulsation modes. On the other hand, Model~3, which produces
a prograde traveling beat pattern (cf.~Fig.~\ref{fig:Si4553_mod90})
does not represent the observations very well and must be excluded.

The weakness of the predicted variations in the H$\alpha$ line as
shown in Fig.~\ref{fig:Halpha_mod09} (displayed with the same
intensity cut levels as the \ion{Si}{iii}$\lambda$4553 line in
Fig.~\ref{fig:Si4553_mod91}) is due to the intrinsically (Stark-)
broadened photospheric H$\alpha$-line profile, which tends to smear
out the NRP-induced line-profile variations over the extended line
profile.  This at first glance surprising result of the NRP modeling
is consistent with the \emph{non-detection} of any of the three NRP
pulsation periods in the observed H$\alpha$ time series.

On the other hand, the finding of NRP beat periods in the order of
several days is of great importance for the investigation of the
possible connections of (fast) photospheric variability on time scales
of hours to (slow) wind variability on time scales of days. The
possible implications will be discussed extensively below in
Sect.~\ref{sect:wind_modeling} where we will try to understand and
model the effects of slowly varying photospheric surface structures on
the stellar wind.

\subsection{The case $l \neq -m$}

\begin{figure}
 \resizebox{\hsize}{!}{\includegraphics[angle=0]{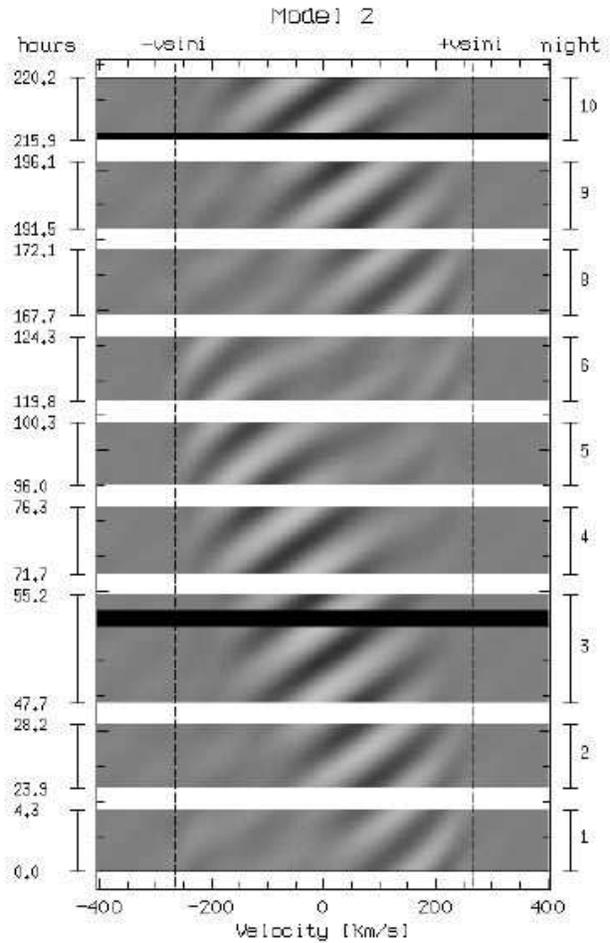}}
 \caption{Computed time series for the line \ion{Si}{iii}$\lambda$4553
   according to Model~2. Note the strong changes from night to night,
   which are due to the superposition of modes with different $l$
   values. Displayed with 1\% cut levels.}
 \label{fig:Si4553_mod92}
\end{figure}

The observed periods $P_1, P_2, P_3$ are closely spaced in time in the
observer's reference frame. However, since the transformation of the
periods to the co-rotating reference frame 
\begin{eqnarray}
 \label{eqn:Pcorot}
  \frac{1}{P_\mathrm{corot}(l,m)} 
      = \frac{1}{P_\mathrm{obs}(l,m)} + m\frac{1}{P_\mathrm{rot}}
\end{eqnarray}
only depends on the pulsation quantum number $m$, the pulsation
periods are actually very different in the co-rotating frame for the
case $l = -m$ as discussed above in Model~1.  The observed close
spacing of the observed periods would be in this case merely an
unlikely coincidence.

Therefore, we tested with Model~2 the case of $l \neq -m$ but with $m$
identical for all three pulsation modes since in this case the close
spacing in the observer's reference frame is naturally explained by a
close spacing in the co-rotating reference frame.

In order to maintain the interesting beat patterns, which appear to be
essential to explain the observed NRPs in the photospheric lines, we
have to assume different $l$-values for the individual modes. In
Fig.~\ref{fig:Si4553_mod92} we show the resulting line profile
variations for one such case, with $l$-values of $10$, $8$ and $6$ and
an $m$-value of $-6$ (Model~2 in Tab.~\ref{tab:brucekylie_modpara2}).
It can be seen that qualitatively similar changes can be produced as
for the case of $l = -m$ and different $l, m$-values. With slightly
larger velocity amplitudes of the pulsations the same amplitude of
profile variations and the same power distribution over the line
profile is produced. The increased velocity amplitudes are needed to
compensate for the cancellation effects from surface zones of opposite
phase. However, despite the fact that the two investigated cases with
properly chosen model parameters lead to practically indistinguishable
phase and power spectra, the actual resulting surface patterns are
clearly more complex in the $l \neq -m$ case and in particular display
strong differences between pole and equator within the same meridional
sector of the star.

\subsection{Multi-line analysis}
\label{sect:multiline}

One possibility to discriminate between $l=|m|$ and $l\neq|m|$ pulsation
modes is a multi-line analysis of the NRPs. The strong variations of
temperature and gravity from pole to equator in a fast rotating and
pulsating star like \object{HD\,64760} provide additional diagnostics by the
analysis of spectral lines with different temperature and gravity
sensitivity. With polar/equator temperatures of 29\,000/23\,300\,K,
the temperature sensitive \ion{He}{ii}$\lambda 4686$ line is of
particular interest in this case.

\begin{figure}
\resizebox{\hsize}{!}{\includegraphics[angle=0]{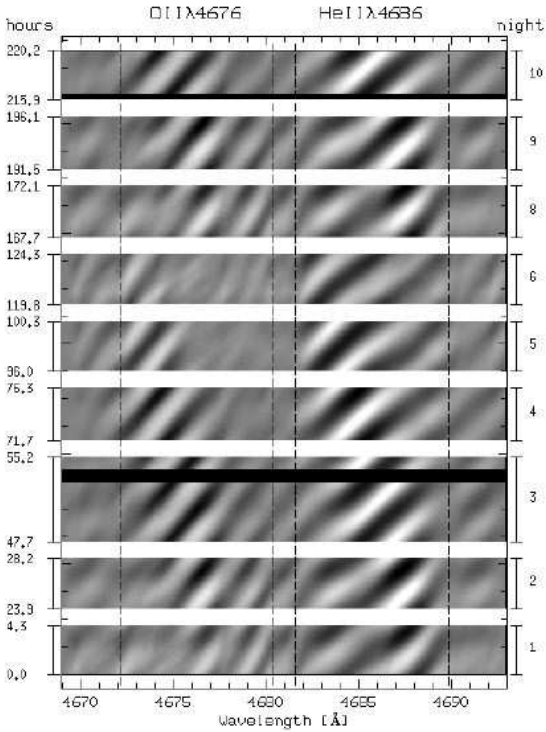}}
\caption{Dynamical spectrum of the \ion{He}{ii}, \ion{O}{ii} spectral
  region computed according to the NRP Model~2 assuming adiabatic
  temperature effects (cf. observed data in
  Fig.~\ref{fig:HeIIOII_ts}). $\pm v\sin i$ for the \ion{O}{ii} (left)
  and the \ion{He}{ii} (right) lines are indicated.}
\label{fig:HeIIOII_model}
\end{figure}

Figure~\ref{fig:HeIIOII_model} shows the synthetic profiles for the
the \ion{He}{ii} line and the neighboring \ion{O}{ii}$\lambda 4676$
line according to the parameters of our $l\neq|m|$ NRP Model 2. While
the global appearance of the two lines is the same, distinct
differences can be seen in the details of the NRP patterns, in
particular in the smaller slope and the lesser extent in velocities of
the features (i.e., they do not reach $\pm v\sin i$). These
characteristics are as expected for the \ion{He}{ii} line, which
is predominantly formed closer to the stellar pole at higher
temperatures: (i) the number of features at a given time remains the
same, (ii) $v\sin i$ becomes apparently smaller for that line since at
the polar caps only smaller projected rotation velocities are seen,
(iii) the slope of the traveling features has to become smaller since the
pulsation periods have not changed.

\begin{figure}
  \resizebox{\hsize}{!}{\includegraphics[angle=0]{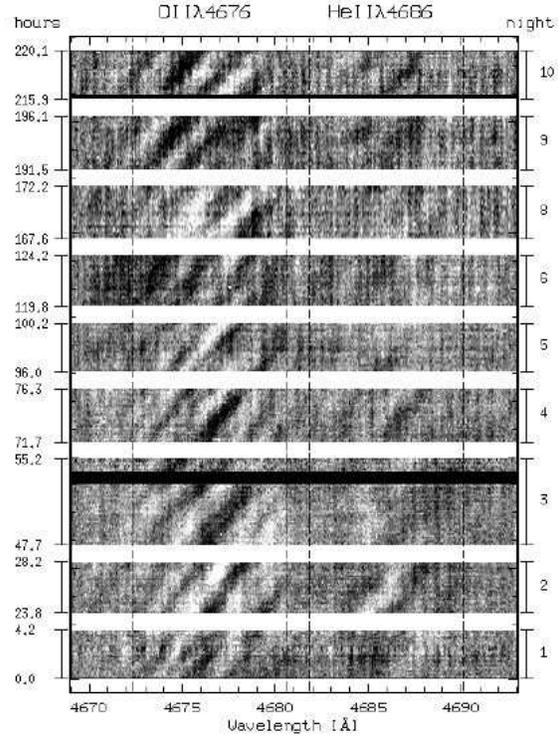}}
\caption{Dynamical spectrum of the {\sc Feros} observations of the
  \ion{He}{ii}, \ion{O}{ii} spectral region. This dynamical spectrum
  has to be compared to the model computations as shown in
  Figs.~\ref{fig:HeIIOII_model} and \ref{fig:HeIIOII_t50_model}. $\pm
  v\sin i$ for the \ion{O}{ii} (left) and the \ion{He}{ii} (right)
  lines are indicated.}
\label{fig:HeIIOII_ts}
\end{figure}

The observed dynamical spectrum as shown in Fig.~\ref{fig:HeIIOII_ts}
corresponds well for the \ion{O}{ii} line as expected due to its
similarities to the \ion{Si}{iii} line for which the model was
constructed before and a good match with the observations was
achieved. However, the model does not match for the \ion{He}{ii} line.
In particular the amplitudes of the variations of the \ion{He}{ii} line
are observed to be very small and hardly discernible with the given
S/N while in the model the amplitudes are predicted to be even larger
than for the \ion{O}{ii} line. 

\begin{figure}
\resizebox{\hsize}{!}{\includegraphics[angle=0]{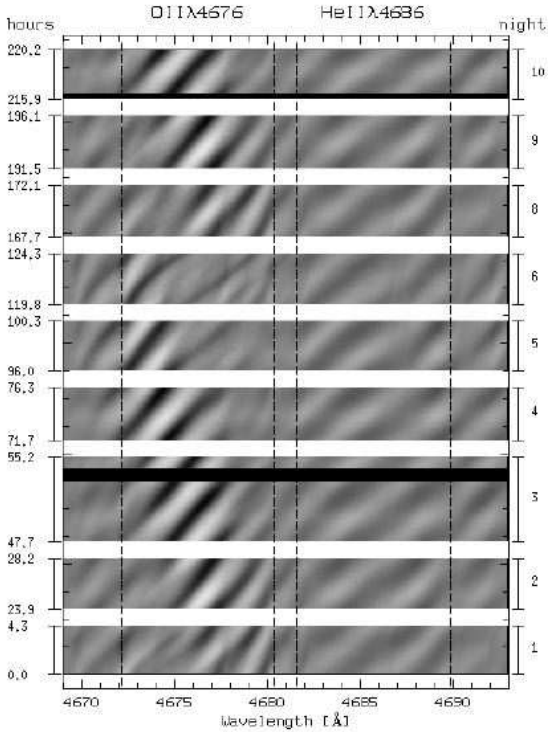}}
\caption{ Dynamical spectrum of the \ion{He}{ii}, \ion{O}{ii} spectral
  region as computed from Model~2 but with the pulsation-induced
  temperature effects reduced to 50\% of the adiabatic value. 
  (cf. observed data in Fig.~\ref{fig:HeIIOII_ts}). $\pm
  v\sin i$ for the \ion{O}{ii} (left) and the \ion{He}{ii} (right)
  lines are indicated.}
\label{fig:HeIIOII_t50_model}
\end{figure}

The fact that the models predict too large variations for the
temperature sensitive \ion{He}{ii} line while less temperature
sensitive lines are well represented indicates that the {\sc Bruce}
models overestimate the pulsation-induced temperature variations. The
models assume adiabatic temperature variations according to
\citet{1979ApJ...232..213B}. A good agreement between observation and
model can be achieved if the temperature effects are reduced to 50\%
of the adiabatic value, which leads to a reduction of the amplitude of
the variations by a factor of 5. The corresponding dynamical spectrum
is shown in Fig.~\ref{fig:HeIIOII_t50_model}. Interestingly,
\citet{1979ApJ...232..213B} found already in their original work on
the non-radial pulsator \object{53\,Persei} that the assumption of
adiabaticity led to an overestimation of the pulsation-induced
photometric variations by one magnitude.  From our observations we
have to conclude that the assumption of adiabaticity also breaks down
for realistic pulsation models of \object{HD\,64760}.

\subsection{Photometry}
\label{sect:photometry}

\begin{figure}
\resizebox{\hsize}{!}{\includegraphics[angle=-90]{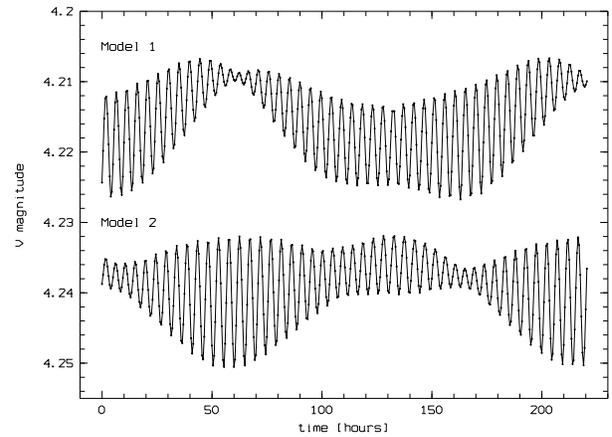}}
\caption{Photometric variations in the Johnson V band as predicted by
  NRP Model~1 (upper light curve) and NRP Model~2 (bottom light curve)
  for the full 10-day observing run and with a time sampling of
  30\,minutes. The magnitude scale of the model has been adjusted to
  match the V magnitude of \object{HD\,64760} from the literature
  ($V=4.24$, \citet{1982bsc..book.....H}).  The synthetic lightcurve
  for Model~1 has been shifted by $-0.025$\,mag for better display. }
\label{fig:nrpphot_model}
\end{figure}
 
Another possibility to discriminate between $l=|m|$ and $l\neq|m|$
pulsation modes is to compare the observed photometric variations with
the photometric variations as predicted from NRP models.

Figure~\ref{fig:nrpphot_model} shows the results from our {\sc
Bruce/Kylie} NRP simulations for the Johnson V filter for the two
studied cases of NRP modes (i.e., Model~1 and 2 as defined in
Tab.~\ref{tab:brucekylie_modpara2}). For both cases, the maximum
(peak-to-peak) amplitude of the variations is 20\,mmag which is
consistent with the photometric variations of up to 30\,mmag as
frequently found in OB supergiants \citep{2004MNRAS.351..552M}.  
the temperature effects in the NRP models are reduced to 50\% of the
adiabatic value as suggested by our analysis in
Sect.~\ref{sect:multiline}, this leads to a reduction of the maximum
amplitude of the photometric variations to 12\,mmag.
  
While the high-frequency variations in the light curves are directly
related to the dominant individual NRP modes with periods close to
4.8\,hours, the overall evolution and the shape of the envelope of the
light curves is governed by the beating of the pulsation modes.  In
particular the clearly different characteristics of the light curves
at times of constructive ($t\approx60, 220$\,hrs) and destructive
interference ($t\approx160$\,hrs) is of high potential to discriminate
between the two studied cases: while in the case of $l=|m|$ (Model~1)
the amplitude of variations is minimal at times of constructive
interference, the amplitude is maximal in the case of $l\neq|m|$
(Model~2) and vice versa at times of destructive
interference. Further, while Model~1 predicts an increase of the
average brightness by 10\,mmag at times of constructive interference,
the variations of the average brightness according to Model~2 are
reduced to some 5\,mmag without clear relation to the evolution of the
beat pattern.
  
{\sc Hipparcos} photometry \citep{1997A&A...323L..61V} of
\object{HD\,64760} (151 measurements over 550\,days) displays
dispersions of 7, 26, and 22\,mmag in the $H_P$, $B_T$ and $V_T$
passbands, respectively. The measured variations are comparable with
our model predictions.  However, no significant periodic signals can
be found in the data sets.  Unfortunately -- to the authors' knowledge
-- no other suited photometric data sets are available for
\object{HD\,64760} to verify the predictions of the NRP models. The
required photometric accuracy of a few mmags appears technically
feasible for a dedicated photometric monitoring campaign, which could
put additional rigid constraints on the NRP modeling, in particular on
the selection of the pulsation mode parameters.

%
%
\section{Stellar-wind modeling}
\label{sect:wind_modeling}

One of the key questions for understanding the nature and origin of
the wind variability in hot stars is its connection to the stellar
photosphere. It is widely assumed, that the winds are modulated by a
mechanism related to the photospheric rotation, presumably patches on
the stellar surfaces produced either by non-radial pulsation (NRP)
patterns or magnetic surface structures ("spots"). The spatial
structures in the photospheres locally change the lower boundary
condition of the stellar wind, which in turn causes localized
structural changes in the wind.  These structural changes modulate the
observed stellar wind profiles as they are dragged through the line of
sight to a distant observer by underlying rotation of the star.
  
In KPS\,2002 we have introduced a simple 1.5D-model (dubbed {\sc
  Rotbalmer}) for a differentially rotating stellar wind to model the
double emission H$\alpha$ profile of \object{HD\,64760}. The model is
based on the work and description of \citet{1996A&A...312..195P} and
allows a correct handling of the rotationally twisted resonance zones
in the line formation of wind lines like H$\alpha$. The model makes
the simplified assumption of a spherically symmetric density
stratification throughout the wind.

To estimate the effects of a geometrically structured rotating wind, we
have introduced in the wind model density enhancements with respect to
the local ambient wind following a streak-line geometry as described
by \citet{1997A&A...327..699F}. The streak line is characterized by the
density contrast $\rho_c$ with respect to the ambient wind, the
angular extension $\phi_s$ on the stellar surface and the rotation
velocity $v_\mathrm{s}$ of the foot point. Usually, $v_\mathrm{s}$
is set equal to the rotation velocity $v_\mathrm{rot}$ of the star
corresponding to a streak line emerging from the stellar surface and
therefore locked to the stellar rotation.

\subsection{The case $v_\mathrm{s} = v_\mathrm{rot}$}

\begin{table}
\caption{{\sc Rotbalmer} model parameters}
\label{tab:rotbalmer_modpara}
\centering
\begin{tabular}{lrl}\hline\hline
\multicolumn{3}{c}{Model Parameters}    \\
\hline
 $T_\mathrm{eff}$      & 24\,600\,K         & effective temperature \\
 $R_\mathrm{equator}$  & 21.6\,R$_\odot$    & radius at equator \\
 $v_\mathrm{equator}$  & 265\,km\,s$^{-1}$  & velocity at equator \\
 $i$                   & 90\,deg            & inclination \\
 $v_\mathrm{inf}$      & 1500\,km\,s$^{-1}$ & wind terminal velocity \\
 $v_\mathrm{min}$      & 10\,km\,s$^{-1}$   & wind start velocity \\
 $\beta$               & 0.8                & wind acceleration parameter \\
 $\dot{M}$             & $9\times10^{-7}$\,M$_\odot$/yr & mass-loss rate \\
 Y$_\mathrm{He}$       & 0.10               & helium abundance n(He)/n(H) \\
 I$_\mathrm{He}$       & 2                  & free electrons per Helium atom \\
\hline
\multicolumn{3}{c}{Streak-line Parameters} \\ 
\multicolumn{3}{c}{Model~1} \\
\hline
 $n_\mathrm{s}$        & 2              &  Number of streak lines \\ 
 $\rho_\mathrm{c}$     & 1.3            &  Density contrast w.r.t. ambient\\
 $\phi_\mathrm{s}$     & 50\,deg        &  Angular extension ($2\arctan{0.5}$)\\
 $v_\mathrm{s}$        & $265$\,km\,s$^{-1}$  & Foot-point rotation velocity \\
\hline  
\multicolumn{3}{c}{Model~2} \\
\hline
 $n_\mathrm{s}$        & 2              &  Number of streak lines \\ 
 $\rho_\mathrm{c}$     & 1.3            &  Density contrast w.r.t. ambient\\
 $\phi_\mathrm{s}$     & 50\,deg        &  Angular extension ($2\arctan{0.5}$)\\
 $v_\mathrm{s}$        & $-80.5$\,km\,s$^{-1}$  & Foot-point rotation velocity \\
\hline
\end{tabular}
\end{table}

\begin{figure}
 \resizebox{\hsize}{!}{\includegraphics[angle=0]{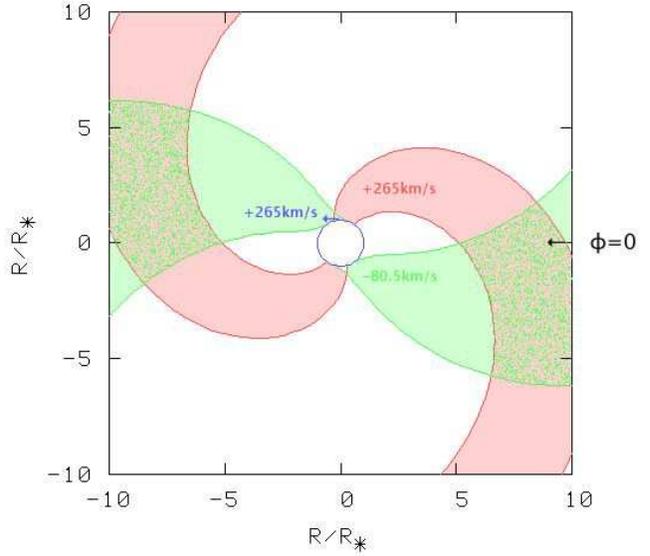}}
 \caption{Streak-line geometry 
            for $v_\mathrm{rot} = +265$\,km\,s$^{-1}$ 
            and $v_\mathrm{s} = +265$\,km\,s$^{-1}$ (Model~1) 
            and $v_\mathrm{rot} = +265$\,km\,s$^{-1}$ 
            and $v_\mathrm{s} = -80.5$\,km\,s$^{-1}$ (Model~2).
            The observer looks at the star from the right at model 
            phase $\phi = 0$.
 }
\label{fig:Halpha_streakline}
\end{figure}

The most natural clock for a strictly periodic wind modulation is
provided by the stellar rotation. For \object{HD\,64760} the upper limit for
the stellar rotation period of $4.2$\,days is within the expected
errors close to twice the $2.4$-day modulation period. If this
estimate is correct, the modulating structure on the stellar surface
must be also persistent over such a time scale and geometrically
divide the stellar surface in two structures with $180$\,degree
azimuthal separation.

Our first {\sc Rotbalmer} wind test model (Model~1)
tries to simulate this situation with the presence of $n_\mathrm{s}=2$
diametrically opposed streak lines in the equatorial plane and with
their foot points rotating at the same speed as the stellar surface,
i.e., $v_\mathrm{s} = v_\mathrm{rot} := v_\mathrm{equator}\sin i$.
The complete set of {\sc Rotbalmer} model parameters used in the model
is given in Tab.~\ref{tab:rotbalmer_modpara} and is basically
identical to the parameter set used in KPS\,2002 except for the
additional parameters to describe the streak lines.  The geometry of
the emerging streak lines is shown in
Fig.~\ref{fig:Halpha_streakline}.


\begin{figure}
 \resizebox{\hsize}{!}{\includegraphics[angle=0]{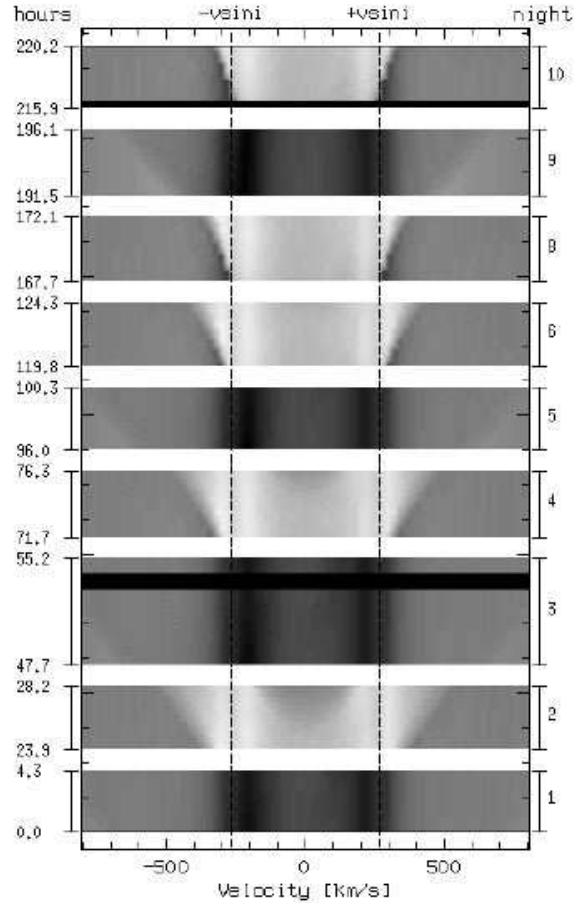}}
\caption{H$\alpha$ {\sc Rotbalmer} Model~1 with 
  $v_\mathrm{s} = v_\mathrm{rot} = 265$\,km\,s$^{-1}$.}
\label{fig:Halpha_model_11}
\end{figure}

Figure~\ref{fig:Halpha_model_11} shows the dynamical H$\alpha$
spectrum as produced by this model with two co-rotating streak lines.
We have again computed the model spectra with the same timings as the
observations to ease their comparison. If the model is compared to the
observations in Fig.~\ref{fig:Halpha_ts} it is found that neither the
velocity evolution of the pseudo-emission and absorption features (the
'X' or '8' shaped pattern) is well represented nor the time scales of
the variations.
The dominant time scale is due to the presence of two streak lines
equal to half the stellar rotation period, i.e., $4.2 / 2 = 2.1$\,days
-- the model had been designed to exactly match this condition. The
model further predicts H$\alpha$ line-profile variations with blue-
and redwards accelerating features as expected for two symmetric
streak lines rotating through the line of sight of the observer.  The
features are seen from zero velocity up to the terminal velocities of
the wind.
This result is comparable to the \object{$\zeta$\,Puppis} (O4\,I(n)f)
models by \citet{2000MNRAS.315..722H} who investigated the effects on
the H$\alpha$ line profile of a narrow and high-contrast one-armed
spiral in the rotating wind of this prototypical O-type star. Indeed,
we have used the model by \citet{2000MNRAS.315..722H} to verify the
functionality of our simpler {\sc Rotbalmer} code and can report that
we find a good agreement in the modeled LPVs.

\subsection{The case $v_\mathrm{s} \neq v_\mathrm{rot}$}

\begin{figure}
 \resizebox{\hsize}{!}{\includegraphics[angle=0]{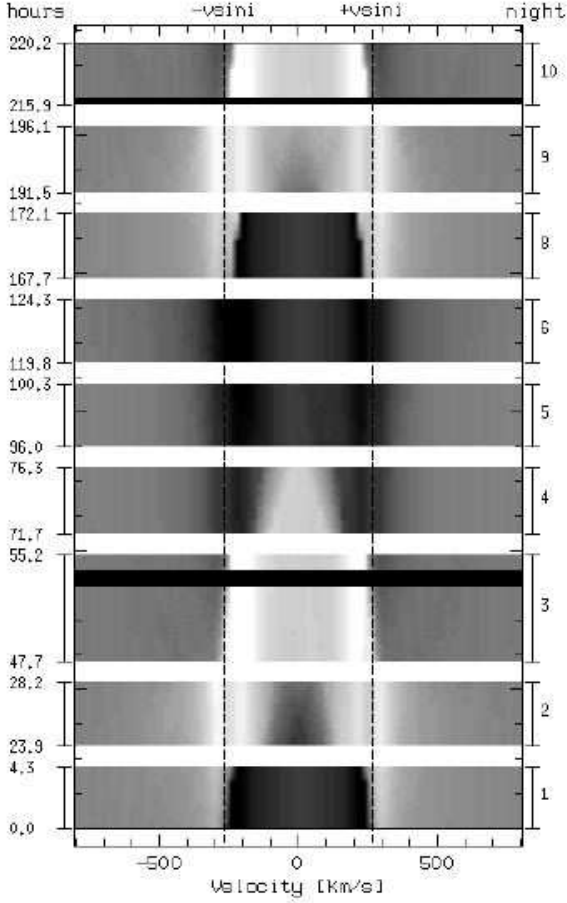}}
\caption{H$\alpha$ {\sc Rotbalmer} Model~2 with
  $v_\mathrm{s} = -80.5$\,km\,s$^{-1}$ corresponding to a beat period
  of 162.8\,hrs.}
\label{fig:Halpha_model_10}
\end{figure}

Our second H$\alpha$ rotating structured wind model assumes again
streak lines spaced at equal angles along the equator.  However, in
contrast to the previous case, we do not assume that the foot points
of the streak lines are fixed on the star and follow its rotation. In
contrast, we assume that the streak lines follow the pulsation
pattern, as resulting from the superposition of the three detected NRP
modes. This pattern does not follow the rotation, but has a pattern
speed slower than the rotation or -- as in our case -- can even move
(in the observer's reference frame) in the opposite direction as the
rotation.  While this does not change the stream line of an individual
particle, the streak lines can change significantly (cf. below).
The number of streak lines depends on the pulsation pattern. In the
case of $l = -m$, two interfering pattern with different $m$ produce
$m_1 - m_2$ maxima along the equator. If $l \neq -m$ and $m$ equal for
both modes, $m$ maxima along the equator appear.  For arbitrary values
of $l$ and $m$ (or more than two interfering modes), the interference
pattern can become very complex. Therefore, we investigate here the
case $l = -m$ (corresponding to our NRP Model 1), which is more
attractive as lower boundary condition for a wind model with
equatorial streak lines because it favors larger pulsation (beat-)
amplitudes at the stellar equator. This second H$\alpha$ rotating
structured wind model assumes again $n_\mathrm{s}=2$ streak lines
diametrically opposite at the equator corresponding to the 2 maxima in
the beat amplitudes as created for $m_1 - m_2 = -2$ .

The period in which the beat pattern moves across the star is related
to the beat period. Therefore, periods significantly longer than the
individual pulsation periods can result. The beat period is not
related to the rotation period of the star, but only to the pulsation
periods. If we consider the two strongest pulsation modes $P_1$ and
$P_2$ only, we obtain ${P_\mathrm{beat}} = 162.8$\,hr
(cf. Eqn.~\ref{eqn:Pbeat}).
The apparent rotation velocity of the maxima of the beat pattern is then 
computed as:
\begin{eqnarray*}
v_\mathrm{s} 
& = & v_\mathrm{rot}  (m_1-m_2) \frac{P_\mathrm{rot}}{P_\mathrm{beat}} \\
& = & 265\,\mathrm{km\,s}^{-1} (-8 - -6) \frac{98.9\,\mathrm{hr}}{162.8\,\mathrm{hr}} \\
 & = & -80.5\,\mathrm{km\,s}^{-1}
\end{eqnarray*}
The geometry of the streak line changes considerably compared to the
co-rotating case as is illustrated in
Fig.~\ref{fig:Halpha_streakline}.  The stream line of the individual
particle ejected from the stellar surface does not change and is ruled
by the $\beta$-law acceleration with the ambient wind in radial and by
the conservation of angular momentum in azimuthal direction. However,
since the spot from where the particles are ejected either rotates
fixed relative to the stellar surface ($v_\mathrm{rot} = v_\mathrm{s}
= +265$\,km\,s$^{-1}$) (Model~1) or moves in the opposite direction of
the rotation ($v_\mathrm{rot} = +265$\,km\,s$^{-1}$ and $v_\mathrm{s}
= -80.5$\,km\,s$^{-1}$) (Model~2) the resulting streak
line, i.e., the snapshot of all ejected particles for a given time,
appears substantially different
(cf.~Fig.~\ref{fig:Halpha_streakline}).

Figure~\ref{fig:Halpha_model_10} shows the correspondingly computed
{\sc Rotbalmer} model (Model~2). Despite being far from perfect, this
model with a counter-rotating pulsation beat pattern and two streak
lines emerging from the maxima of beat amplitudes does show some
resemblance with the observed '8' or 'X'-shaped pattern in the
dynamical spectra in Fig.~\ref{fig:Halpha_ts}. Particularly remarkable
is the evolution of the pseudo-emission features from wind velocities
beyond $\pm v\sin i$ to zero-velocity in nights 1 to 4 and the phase
of no variability in night 5, which are mostly responsible for the
resemblance with the observed 'crossing' features.  Also the lack of
features accelerating with the wind matches the observations better
than the first model.  Most important, also the time scales based on
the beat period and the presence of two maxima in the beat amplitudes
on the stellar equator match the observations surprisingly well.
Clearly, the observed pseudo-emission and absorption features are more
'discrete' in velocity and time than the model prediction. But
considering the large simplifications and the complete lack of physics
in the transition zone from the pulsating photosphere to the base of
the wind in our model, the result is encouraging for further detailed
modeling.

The important finding from our simplistic model therefore is, that the
constructive interference of multi-periodic NRPs could play the
crucial rule of providing the surface structure moving with the right
time scales at the base of the wind to match (i) the observed
base-of-the-wind variability (H$\alpha$) and (ii) the outer-wind
structures rotating through the observer's line of sight as observed
in the UV.

%
%
\section{Discussion}

\begin{figure}
 \resizebox{\hsize}{!}{\includegraphics[angle=0]{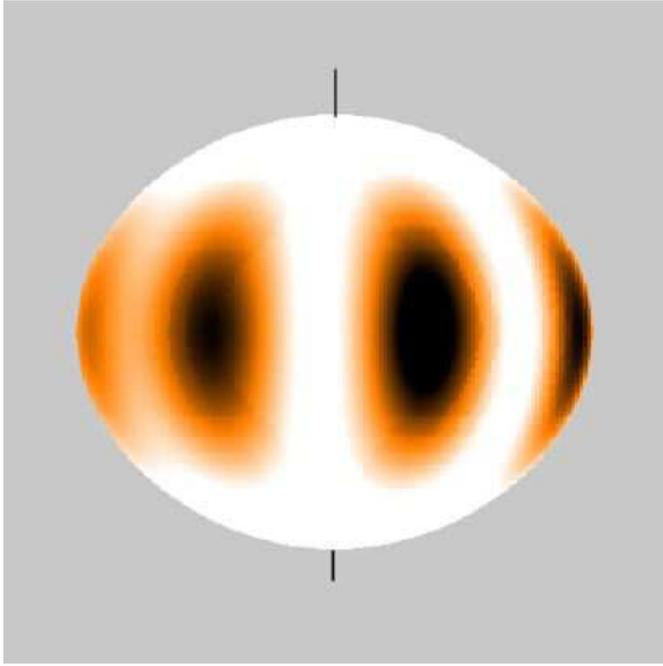}}
\caption{Temperature distribution on the stellar surface due to NRP
  beat patterns (Model~1) as seen from the equatorial plane at
  $t=48$\,hrs. The minimum and maximum temperatures in the visible
  regions are $T=21\,500$\,K (black) and $T=25\,000$\,K (white).}
\label{fig:NRP_mod08_teff}
\end{figure}

The new high-resolution optical times series presented in this work
allowed us to further investigate the line-profile variability of the
B-supergiant \object{HD\,64760} in the photosphere and at the base of
the wind. This new {\sc Feros} data set provides detailed information
to extend the {\sc Heros} data set (KPS\,2002).

The high quality and the good sampling of the photospheric and wind
variability over more than 2 rotational cycles of the star allowed us
to reveal the presence of multi-periodic line-profile variations
(LPVs) in the stellar photosphere. The three closely spaced periods
could be identified as prograde high-order non-radial pulsation (NRP)
modes with $l$ and $|m|$ values in the range $6 - 8$. The first
detection of high-order NRPs at a period of $~0.1$\,days had been
reported by \citet{1984LIACo..25..115B} -- however, his data set was
not adequate for a possible detection of closely spaced multi-periodic
NRPs. Nevertheless, even the quality of our data set did not allow a
very precise determination of the three pulsation modes from the
small-amplitude ($<0.5$\%) LPVs. Therefore, we tested the two
fundamental cases $l=-m$ and $l\neq-m$ via detailed modeling of the
multi-periodic LPVs. Both cases allow us to represent the observations
equally well. However, we favor the case $l\neq-m$ to account for the
fact that the observed 3\% splitting into the three periods
($P_1=4.810$\,hr, $P_2=4.672$\,hr, $P_3=4.967$\,hr) is a true physical
and asteroseismologically significant period splitting in the star's
co-rotating frame, which implies that the $m$-values of the three
modes have to be equal and different from $l$-values (our Model~2,
cf.~Tab.~\ref{tab:brucekylie_modpara2}). 
In the case of rotational splitting of pulsation modes, the expected
magnitude of the effect is such that the beating of rotationally split
frequencies is of the order of the rotational frequency. Considering
the measured split frequencies, it is found that their beat frequency
is roughly consistent with the estimated rotational frequency of
\object{HD\,64760}. However, additional complications to this simple
estimate have to be considered for an asteroseismological application
of measured rotational splitting \citep{1999ApJ...516..349K}.
It should be kept in mind that -- if our preference for the $l\neq-m$
case is correct -- we are facing highly complex NRP surface patterns
in \object{HD\,64760}. For this reason we were restricted to use the
geometrically simpler $l=-m$ model for our subsequent modeling of
stellar wind structure rooted in photospheric NRPs.

The shortness of the measured periods and the absence of significant
horizontal velocity fields clearly identifies the pulsation modes as
p-modes. The pulsation parameters with periods of a few hours as found
for \object{HD\,64760} compare well with the findings for other O-type
stars (cf. e.g. \citet{1999LNP...523..305H} for a compilation).
It is particularly interesting that the first NRPs in early-type
supergiants were discovered in \object{$\gamma$\,Arae} (B1\.1b) by
\citet{1984A&A...140...72B} with \object{$\gamma$\,Arae} being almost
a twin to \object{HD\,64760} with respect to the stellar parameters,
including the very high value of $v\sin i$. In particular the
short-term variability in the photospheric lines of
\object{$\gamma$\,Arae} with a derived period of 0.17\,days and $l=10$
is very reminiscent of the variability seen in this work in
\object{HD\,64760}. However, it is important to note that the
variability in the UV wind-lines of \object{$\gamma$\,Arae} is quite
distinct from the periodic behavior of \object{HD\,64760},
cf. \citet{2002A&A...388..587P}.

\citet{1997A&A...327..699F} had proposed as one possible source of
regular surface structure in \object{HD\,64760} NRP patterns, which
are locked to the 4.8-day rotation of the star, i.e., low-order
g-modes with observed pulsation periods of 2.4 and 1.2\,days and
$l=-m$-values of 2 and 4, respectively.  However, this scenario would
imply an infinitely long pulsation period in the co-rotating frame
(cf. Eqn.~\ref{eqn:Pcorot}), which appears not to be very
physical. Further, we have no evidence from models of pulsational
instabilities for very long pulsation periods, for which the rotation
could become the dominant time scale in the observer's frame.

KPS\,2002 attributed the observed 2.4-day period in the H$\alpha$
variability to modulations of the inner flanks of the emission humps
at photospheric velocities to complex \emph{width} variations of the
underlying photospheric H$\alpha$ profile and tried to connect these
variations directly to the effect of NRP-induced LPVs of the
photospheric H$\alpha$ profile. The fact that the strongest
variability in the line profile is seen close to velocities of $\pm
v\sin i$ was taken as hint for the presence of strong horizontal
velocity fields, i.e., the presence of low-order g-mode
pulsations. The marginal detection of NRP-like LPVs in the
\ion{He}{i}$\lambda4026$ line further supported this hypothesis.  With
the new 2003 data set and the modeling results as presented in this
work, this picture has to be refined. As shown in
Sect.~\ref{sect:nrp_modeling}, NRPs are not expected to appear
directly as measurable LPVs of the photospheric H$\alpha$ profile due
to smearing of the variations over the line profile as caused by the
Stark-broadening of the intrinsic line profiles. Further, the new data
set could not confirm the existence of NRP-induced LPVs on time scales
of the wind modulation period but revealed the presence of
multi-periodic NRPs, which are identified as p-mode pulsations with
small horizontal velocity fields and periods of hours.  Therefore, the
observed H$\alpha$ wind variations on time scales of days can only be
attributed to \emph{secondary} effects possibly linked to the presence
of NRPs as discussed below.  The interpretation of the observed
H$\alpha$ wind variations is further complicated by the presence of
the differentially rotating wind: our simulations with the {\sc
Rotbalmer} code indicate that a simple direct mapping of velocities in
the line profile to a location in the stellar wind is not possible in
the presence of rotationally twisted resonance zones. Indeed, any
large-scale wind variability (like rotating streak lines) is primarily
mapped into LPVs close to $\pm v\sin i$.  The power distribution over
the line profile as e.g.  derived from the models shown in
Figs.~\ref{fig:Halpha_model_11} and \ref{fig:Halpha_model_10} display
four power peaks blue- and red-wards of $\pm v\sin i$ which nicely
reproduces the observed power distribution of the observed H$\alpha$
variability (cf.  KPS\,2002, their Fig.~3, blue and red peaks 'A' and
'D').

The most striking consequence from our finding of multi-mode NRPs with
periods of a few hours is the presence of beat periods of several
days. These beat periods can be directly seen in the observed and
modeled photospheric LPVs as \emph{retrograde} patterns traveling
slowly (days) over the line profile superimposed on the fast (hours)
\emph{prograde} patterns of the individual pulsation modes (see
e.g.~Fig.~\ref{fig:Si4553_mod91}). Simultaneously, we find the
variability of the mostly wind-sensitive H$\alpha$ line to be of the
same time scales. Even some characteristics of the H$\alpha$
line-profile variability like phases of low wind variability seem to
coincide with phases of small beat amplitudes in the photospheric NRPs
(cf.~Fig.~\ref{fig:Halpha_ts}). The \ion{He}{i}$\lambda$6678 line
(cf.~Fig.~\ref{fig:He6678_ts}), which shows on top of the photospheric
NRP patterns some wind emission features, demonstrates directly the
connection of the retrograde traveling NRP beat pattern with the wind
emission, which apparently follows the beat pattern over the line
profile (notably over the days 6-9). On the other hand, the NRP models
of H$\alpha$ have shown us that the photospheric NRPs themselves do
\emph{not} contribute to the LPVs of H$\alpha$ 
(cf.~Fig.~\ref{fig:Halpha_mod09}).

This is obviously an exciting finding since it provides us for first
time with direct observational evidence for a connection of
photospheric multi-periodic NRPs with spatially structured winds. The
constructive or destructive interference of multiple photospheric
pulsation modes with time scales of a few hours (the intrinsic stellar
pulsation time scale of \object{HD\,64760} is $<1$\,day if a pulsation
constant of $Q=0.04$\,days \citep{1984A&A...133..307L} and the stellar
parameters in Tab.~\ref{tab:brucekylie_modpara1} are assumed) allows
us to connect the fast photospheric variations at the base of the wind
with the long wind modulation periods of days.
Admittedly, the derived beat period of 6.8\,days, which is observed in
photosphere and base of the wind does not directly match the 1.2 and
2.4-day periods, which were derived for the periodic wind variability,
but is more consistent with the longer 5.5 or 11-day repetition time
scales as seen for the slow discrete absorption components (DACs) in
the {\sc Iue Mega} campaign data sets.  
In particular the simultaneous presence of the 2.4-day period on one
side and the wind emission features following the slower beat pattern
on the other side and the lack of an obvious link between the two
periodicities inhibits a simple unified picture of the photosphere --
wind connection in \object{HD\,64760}. Therefore, it appears
particularly important to better understand the actual physical
processes, which can connect photosphere structure to stellar wind
structure. Without such an understanding it will not be possible to
assess the significance of any observed and derived photosphere- and
wind-variability periods. 

Figure~\ref{fig:NRP_mod08_teff} illustrates the spatial temperature
structures in the photosphere created by the beating of the dominant
two periods in our NRP Model~1. The model predicts two diametrically
opposite regions of maximum beat amplitudes (constructive
interference) with temperatures ranging from $T=21\,500$\,K to
$T=25\,000$\,K at the stellar equator. In the regions of destructive
interference $90$\,degrees away in azimuth, the temperature amplitudes
are reduced and range from $22\,300$\,K to $23\,800$\,K. In our
simplified scenario these NRP-induced spatial structures locally
change the lower boundary conditions of the stellar wind, which in
turn causes localized structural changes in the wind. (Note, that we
used a simple 2 bright -- 2 dark spot geometry to represent the
regions of constructive and destructive interference of NRP modes).

The creation of wind structure by variation of lower boundary
conditions have been investigated with hydrodynamical simulations by
\citet{1996ApJ...462..469C} who confirmed the formation of spiral
structures (co-rotating interaction regions, CIR) originating from the
collision of fast and slow streams rooted in the stellar surface.
These wind structures modulate the observed stellar wind profiles as
they are dragged through the line of sight to a distant observer by
the underlying rotation of the stellar wind.

In contrast to earlier models for the localized regions on the stellar
surface of \object{HD\,64760}, we investigated the effects of the
observed and predicted \emph{retrograde} motion of the regions of
maximum beat amplitudes (equal to the regions causing the variation of
the lower wind boundary condition) on the geometry of the spiral
structures and the resulting H$\alpha$ LPVs. Despite the fact that our
H$\alpha$ simplistic model does not include any hydrodynamics but is
mostly geometrical, some key characteristics of the observed H$\alpha$
variations could be reproduced to some extent, which was \emph{not}
possible with models with the surface structures locked to the stellar
rotation: the relevant time scales, and the velocity evolution of the
pseudo-emission and absorption features.

The crucial question to be answered is whether the NRP-induced
variations of the physical parameters like local temperature, gravity,
and velocity fields are actually sufficiently large to introduce
significant variations at the base of the wind to build up wind
structures.
\citet{1996ApJ...462..469C} define the spot brightness amplitude $A$
(with their $T_0$ corresponding to our $T_\mathrm{equator}$) as:
\begin{eqnarray*}
A & = & (T_\mathrm{spot}^4 - T_\mathrm{equator}^4) / T_\mathrm{equator}^4
\end{eqnarray*}
Using the 'spot' temperatures from our NRP Model~1, we obtain amplitudes of 
\begin{eqnarray*}
A^+  & = & ((25\,000\,\mathrm{K})^4 - (23\,300\,\mathrm{K})^4) / 
            (23\,300\,\mathrm{K})^4 \\
     & = & +0.325 \\
A^-  & = & ((21\,500\,\mathrm{K})^4 - (23\,300\,\mathrm{K})^4) / 
            (23\,300\,\mathrm{K})^4 \\
     & = & -0.275 
\end{eqnarray*}
which are comparable to the $A=\pm0.5$ standard spot model used by
\citet{1996ApJ...462..469C} to model the {\sc Iue} wind profile
modulations of \object{HD\,64760} with CIRs. The azimuthal extension
of the standard spot model is with $\phi=20$\,deg comparable to
the azimuthal extension of the bright and dark sectors of our NRP
Model~1.
For the same model but with the temperature effects reduced to 50\% to
account for non-adiabatic effects (cf. Sect.~\ref{sect:multiline}) we
obtain spot brightness amplitudes of $A^+=0.164$ and $A^-=-0.146$ --
according to \citet{1996ApJ...462..469C} still sufficiently large to
yield visible wind structure.

While NRPs are not obviously very efficient in providing localized
bright spots on the stellar surface for ``Radiational Driven Orbital
Mass Ejection'' (RDOME) or ``Pulsational Driven Orbital Mass
Ejection'' (PDOME) in the context of the creation of Be-star disks
\citep{2004IAUS..215..515O}, they might be the source of the
structured winds of hot supergiants like \object{HD\,64760} as we have
tried to demonstrate in this work.

However, a hydrodynamical modeling, which takes fully into account the
geometrical and physical boundary conditions as created by
multi-periodic NRPs in the photosphere (as e.g. shown for the simple
$l=-m$ case in Fig.~\ref{fig:NRP_mod08_teff}) and which further
includes the retrograde motion of the beat pattern with respect to the
stellar rotation is still urgently needed to provide a consistent and
time dependent picture of the stellar photosphere-to-wind structure of
\object{HD\,64760}.

%
%
\begin{acknowledgements}
OS wants to thank the ESO Visiting Scientist Program for the 
support of this project and the ESO Vitacura Science Office
for its hospitality during a extended visit where a major part 
of this work was done.
\end{acknowledgements}

%
%
\bibliographystyle{aa}
\bibliography{3847}

%
%

%
%
\end{document}